\documentclass[a4paper]{scrartcl}

\usepackage{lmodern}
\usepackage{amssymb,amsmath}
\usepackage[T1]{fontenc}
\usepackage[utf8]{inputenc}
\usepackage{upquote}
\usepackage[tracking=smallcaps,letterspace=70]{microtype}
\usepackage[unicode=true]{hyperref}
\usepackage{longtable,booktabs}
\usepackage{graphicx,grffile}
\usepackage{authblk}
\usepackage[comma,super,sort&compress]{natbib}
\usepackage{tabularx}
\newcolumntype{Y}{>{\centering\arraybackslash}X}
\usepackage[font=small]{caption}
\usepackage{multirow}
\usepackage{subfig}
\usepackage{scalerel}
\usepackage{comment}
\usepackage{cases}

\usepackage[british]{babel}
\usepackage[modulo]{lineno}

\UseMicrotypeSet[protrusion]{basicmath}

\providecommand{\subtitle}[1]{}

\PassOptionsToPackage{usenames,dvipsnames}{color}

\hypersetup{%
	pdftitle={Effects of non-parallelism on standard and magnetorheological measurements},
	pdfauthor={R. Rodrigues; F.J. Galindo-Rosales; L. Campo-Dea{\~n}o},
	pdfborder={0 0 0},
	colorlinks=true,
	linkcolor=Blue,
	citecolor=Blue,
	urlcolor=Blue,
	breaklinks=true
}%

\makeatletter
	\def\maxwidth{\ifdim\Gin@nat@width>\linewidth\linewidth\else\Gin@nat@width\fi}
	\def\maxheight{\ifdim\Gin@nat@height>\textheight\textheight\else\Gin@nat@height\fi}
\makeatother

\setkeys{Gin}{width=\maxwidth,height=\maxheight,keepaspectratio}

\ifx\paragraph\undefined\else
	\let\oldparagraph\paragraph
	\renewcommand{\paragraph}[1]{\oldparagraph{#1}\mbox{}}
\fi
\ifx\subparagraph\undefined\else
	\let\oldsubparagraph\subparagraph
	\renewcommand{\subparagraph}[1]{\oldsubparagraph{#1}\mbox{}}
\fi

\title{Effects of non-parallelism on standard and magnetorheological measurements}
\author[1,3]{R. Rodrigues}
\author[2,3]{F.J. Galindo-Rosales}
\author[1,3$\dagger$]{L. Campo-Dea{\~n}o}

\affil[1]{CEFT - Centro de Estudos de Fen{\'o}menos de Transporte, Depto.~de Engenharia Mec{\^a}nica, Faculdade de Engenharia, Universidade do Porto, Rua Dr.\ Roberto Frias, 4200-465, Porto, Portugal} 
\affil[2]{CEFT - Centro de Estudos de Fen{\'o}menos de Transporte, Depto.~de Engenharia Química, Faculdade de Engenharia, Universidade do Porto, Rua Dr.\ Roberto Frias, 4200-465, Porto, Portugal} 
\affil[3]{ALiCE - Laborat{\'o}rio Associado em Engenharia Qu{\'i}mica, Faculdade de Engenharia, Universidade do Porto, Rua Dr.\ Roberto Frias, 4200-465, Porto, Portugal}
\affil[$\dagger$]{Email: \href{mailto:campo@fe.up.pt}{campo@fe.up.pt}}

\date{}

\begin{document}
%\begin{bibunit}[unsrtnat]

\maketitle

\begin{abstract} % about 150 words
\small
	\noindent Aiming towards the magnetorheological characterisation of whole human blood, we evaluated the suitability of our experimental setup for steady shear measurements with low-viscosity fluids. Previous measurements with a rotational rheometer equipped with a magnetorheological cell returned low and inconsistent apparent-viscosity values. In this work, a parametric study was conducted, experimentally and numerically, to evaluate the possible error sources and define an experimentally reliable window. Steady shear measurements were carried out with Newtonian fluids using two geometries: parallel-plates at different gap heights, and cone-plate. A clear decrease in measured viscosity with parallel-plate gap reduction was found, along with a slight overestimation of the cone-plate. Numerical results corroborated the experimental observations, pointing towards a small inclination of the bottom plate (between 0.1$^\circ$ and 0.3$^\circ$). The numerical study has also shown that geometry non-parallelism can lead to significant flow alterations, particularly for cone-plate geometries where contraction/expansion flow can be generated, critically deviating from the canonical Couette flow and provoking non-negligible errors. Additional experimental and numerical work was conducted to evaluate the effects of the non-parallelism on magnetorheological measurements. The geometrical asymmetry results in severe alterations of the microstructural dynamics, possibly leading to an underestimation of the magnetorheological response. This work has shown that geometry non-parallelism can have critical repercussions on simple shear flows, with fundamentally different implications from a general gap-error, and can be identified through comparative measurements with Newtonian fluids in different geometries (parallel-plates and cone-plate).\vspace{.5cm}
 
\end{abstract}

\newpage

%\setkeys{Gin}{draft}

\section{Introduction}
\label{sec:intro}

Blood is a complex fluid due to its composition and mechanical properties. It is well documented that blood is a viscoelastic shear-thinning fluid, with a relatively low high-shear viscosity\citep{valant2011,campo2013,robertson2008} (around 3 mPa$\cdot$s). Moreover, the iron-containing haemoglobin present in red blood cells also gives the biofluid significant magnetic susceptibility\citep{zborowski2003,tao2011}. In recent years, the scientific community has intensified its efforts to understand the effects of magnetic fields on blood's mechanical properties. Magnetic fields have been studied as means of guidance for microrobots and particles for a multitude of medical applications \citep{khalil2014,fu2015,mathieu2009,shaw2010} and also to knowingly alter blood's properties themselves \citep{tao2011}, making magnetohaemorheology a field of significant interest. However, experimental data is still lacking, and a full magnetorheological description of whole blood has yet to be achieved.

Magnetorheological characterisations are commonly performed using rotational rheometers equipped with electromagnetic circuits (magnetorheological cells) that allow the generation of uniform magnetic fields of varying intensity. However, measurements with low-viscosity fluids, such as blood, can be strenuous. There is a plethora of phenomena and instrumental limitations that may compromise measurement quality and, because whole blood is a precious fluid, it is vital to identify the possible error sources and define an experimental window before tackling its magnetorheological characterisation. As such, in this work we set out to evaluate the suitability of our experimental setup for blood testing, performing standard steady shear measurements with different Newtonian fluids using two static bottom plates and two planar geometries (one parallel-plates and one cone-plate). Measurements were also conducted with a Newtonian blood analogue seeded with paramagnetic particles under the influence of an external magnetic field, testing various field densities and particle concentrations. Numerical simulations were additionally conducted to evaluate the effects of a possible error source on standard and magnetorheological measurements.

Before heading to the conducted work, it is relevant to briefly address some common error sources in rotational rheometry that could negatively affect the results (an in-depth walkthrough of such experimental difficulties is given in the seminal work of \citet{ewoldt2015}). Relevant to this work, in the next subsection we shall discuss the flow of low-viscosity Newtonian fluids under steady shear in planar geometries (parallel-plates and cone-plate) and briefly discuss measurements with suspensions and magnetorheometry.

\subsection{Common errors in steady shear measurements}

In a rotational rheometer, the steady shear viscosity, $\eta$, can be obtained through the shear stress, $\tau$, and the shear rate\footnote{Either can be the controlled input, depending on the instrument.}, $\dot\gamma$:
\begin{equation}
    \eta = \frac{\tau}{\dot\gamma} = \frac{\mathrm{F}_\tau T}{\mathrm{F}_\mathrm{\gamma}\Omega}\,,
    \label{eq:viscosity}
\end{equation}
where $T$ is the torque and $\Omega$ is the rotational velocity. $\mathrm{F}_\tau$ and $\mathrm{F}_\mathrm{\gamma}$ are functions of geometrical characteristics, which, for parallel-plates (PP) and cone-plate (CP) geometries, are given by:
\begin{numcases}{\text{Geometrical functions:}}\;
  \mathrm{PP:}\;\;\;\mathrm{F}_\tau = \frac{2}{\pi R^3}\;\;\;\mathrm{and}\;\;\;\mathrm{F}_\mathrm{\gamma} = \frac{R}{h}\,,\nonumber\\[10pt]
  \mathrm{CP:}\;\;\;\mathrm{F}_\tau = \frac{3}{2 \pi R^3}\;\;\;\mathrm{and}\;\;\;\mathrm{F}_\mathrm{\gamma} = \frac{1}{\beta}\,,\nonumber
\end{numcases}
where $R$ is the geometry radius, $h$ is the gap between parallel-plates and $\beta$ is the CP cone angle\citep{ewoldt2015}.

Considering the rheometer specifications, steady shear measurements are limited by the instrument's minimum and maximum torques, $\mathrm{T}_\mathrm{min}$ and $\mathrm{T}_\mathrm{max}$, and maximum angular velocity, $\Omega_\mathrm{max}$. Given that the viscosity results are usually plotted as a function of the shear rate, the valid range in the $\eta(\dot\gamma)$ space is limited by the following expressions:
\begin{equation*}
    \text{Torque limits:}\;\;\;\frac{\mathrm{F}_\tau \mathrm{T}_\mathrm{min}}{\dot\gamma}<\eta<\frac{\mathrm{F}_\tau \mathrm{T}_\mathrm{max}}{\dot\gamma}\,,
\end{equation*}
\begin{equation*}
   \text{Rotational velocity limit:}\;\;\;\dot\gamma<\mathrm{F}_\mathrm{\gamma}\Omega_\mathrm{max}\,.
\end{equation*}

Viscosity data is usually gathered through shear rate sweeps, implying shear rate steps between data points. The finite acceleration of the rheometer and the non-instantaneous re-stabilisation of the flow may, therefore, lead to data acquisition at an unsteady flow state. Thus, a constant time interval is usually defined between shear rate step and data acquisition. Another approach is to define a shear-rate-dependent time function that establishes this waiting period, which is helpful because, usually, the flow stabilises faster for larger rotational velocities. As such, having the time step diminish with the shear rate may reduce the overall measurement time. In any case, preliminary constant-shear-rate measurements are usually performed to evaluate the required time intervals to guaranty steady flow.

At sufficiently large rotational velocities, centrifugal forces induce radial velocity components, i.e., secondary flows. This phenomenon leads to an increase in measured torque and is responsible for an apparent shear-thickening at high shear rates. From an analysis of experimental data, \citet{turian1972} arrived at an expression that estimates the experimental region free from secondary flows:
\begin{equation*}
    \text{Secondary flow limit:}\;\;\;\eta>\frac{\rho\dot\gamma R^2}{4{\mathrm{F}_\mathrm{\gamma}}^3}\,.
\end{equation*}
Increasing the rotational velocity further, the sample may escape the geometry, leading to an abrupt drop in measured viscosity.

Considering the loading conditions of the sample, over- and underfilling can lead to significant errors. Overfilling the geometry can result in additional torque depending on the amount of extra sample volume and geometrical characteristics \citep{cardinaels2019}. On the other hand, underfilling will result in a lower apparent viscosity because the sample cannot completely fill the geometry. Additionally, the symmetry of the sample/air interface is paramount for low shear measurements. \citet{johnston2013} concluded from a mathematical analysis that an additional torque can arise from non-constant surface tension, or non-constant contact angle, at the fluid/air interface. Essentially, asymmetries in the contact line lead to surface tension forces that pull along the interface, resulting in an additional torque, which is more significant at low shear, thus leading to an apparent shear-thinning behaviour that vanishes at sufficiently high shear rates. Sample evaporation and over/underfilling are also sources of surface tension torque\citep{johnston2013}. This issue can be mitigated by an accurate sample volume, reducing sample evaporation and diminishing the fluid's surface tension \citep{johnston2013}. Moreover, carefully centring the sample in the geometry aids in reducing contact line asymmetry as far as the wetting conditions of the plates allow. Despite all the efforts to counteract this phenomenon, it can still lead to torque artefacts up to two orders of magnitude larger than the rheometer's minimum torque. Thus, it has been common practice to multiply the low-torque limit by a safety coefficient, shifting it to higher shear.

Geometrical characteristics can also affect measurement quality. Deviations from the nominal dimensions and unwanted rougher surface finishing from fabrication errors or wear can result in incorrectly measured viscosities. Apart from the geometry's characteristics, errors in the gap height between the plates (or the cone and plate) can also lead to significant errors. Gap-errors can arise from a multitude of factors \citep{davies2008}. Setting zero-gap is usually done by lowering the moving plate until an increase in normal force is sensed, signalling solid-solid contact. However, if the threshold force for contact assumption is not sufficiently large ($\lesssim5$ N), normal forces arising from air squeeze can incorrectly serve as the contact input, leading to a deviation of the zero-gap by a few micrometers\citep{davies2005}. Deviations in the surface finishing of either plate and non-parallelism of the geometry may also lead to gap errors. In any case, the usual gap-error leads to the actual/average gap in the geometry being larger than the commanded gap, typically around 10 to 50 \textmu m\citep{ewoldt2015}. This can lead to sample underfilling if the loading quality is not, or cannot be verified before initiating the measurement.
Moreover, even with perfect loading, gap-errors must not be disregarded, particularly for small gaps. Considering a PP system (for CP geometries, the gap is fixed by default), having the real gap as the sum of the commanded gap and a gap-error: $h = h_\mathrm{c}+\epsilon$, a divergence between the commanded and applied shear rates arises: $\dot\gamma_\mathrm{c} = (R/h_\mathrm{c})\,\Omega$, $\dot\gamma_\mathrm{r} = (R/(h_\mathrm{c}+\epsilon))\Omega$, which leads to a viscosity underestimation that gains relevance as the gap is decreased. With some manipulation, we can obtain:
\begin{equation*}
    \text{Gap-error formulation:}\;\;\;\frac{h_\mathrm{c}}{\eta_\mathrm{m}} = \left(\frac{1}{\eta_\mathrm{r}}\right)h_\mathrm{c} + \frac{\epsilon}{\eta_\mathrm{r}}\,,
\end{equation*}
where $\eta_\mathrm{m}$ and $\eta_\mathrm{r}$ are the measured and real viscosities, respectively. This expression is a variation of the one given by \citet{kramer1987} and allows for an estimation of the fluid's true viscosity, $\eta_\mathrm{r}$, and the gap-error, $\epsilon$, by conducting measurements at multiple gap heights. It is, however, noteworthy that through this analysis slip effects cannot be decoupled from a gap-error as these return a similar measured-viscosity decrease with gap height reduction\citep{vleminckx2016}.

For future reference, the Equations relating to the discussed experimental limits and the gap-error formulation are surmised in Table \ref{tab:equations}.

\newcommand{\eqnum}{\leavevmode\hfill\refstepcounter{equation}\textup{(\theequation)}}

\begin{table}[htp]
    \center
    \renewcommand{\arraystretch}{2.5}
    \small{
    \caption{Overview of the general equations: geometric functions, experimental limits and gap-error formulation}
    \label{tab:equations}
    \begin{tabularx}{\textwidth}{@{}>{\raggedright\arraybackslash\hsize=.2\hsize}X|>{\raggedright\arraybackslash\hsize=1.5\hsize}X|>{\raggedright\arraybackslash\hsize=0.25\hsize}X>{\centering\hsize=0.9\hsize}X>{\centering\hsize=0.9\hsize}X>{\raggedleft\arraybackslash\hsize=0.25\hsize}X@{}}
    \hline
    \multicolumn{2}{c|}{\multirow{2}{*}{\vspace{-10pt}\centering Geometric functions}}&
    PP&
    $\displaystyle\mathrm{F}_\tau = \frac{2}{\pi R^3}$&
    $\displaystyle\mathrm{F}_\mathrm{\gamma} =\frac{R}{h}$&\eqnum\\[5pt]
    %\hline
    \multicolumn{2}{c|}{}&CP&
    $\displaystyle\mathrm{F}_\tau = \frac{3}{2 \pi R^3}$&
    $\displaystyle\mathrm{F}_\mathrm{\gamma} = \frac{1}{\beta}$&\eqnum\\[5pt]
    \hline
    \multirow{4}{*}{\vspace{-20pt}\rotatebox[origin=c]{90}{Experimental limits}}&
    Minimum torque&&\multicolumn{2}{c}{$\eta>\displaystyle\frac{\mathrm{F}_\tau \mathrm{T}_\mathrm{min}}{\dot\gamma}$}&
    \eqnum\label{eq:min_torque}\\[5pt]
    %\hline
    &Maximum torque&&\multicolumn{2}{c}{$\displaystyle\eta<\frac{\mathrm{F}_\tau \mathrm{T}_\mathrm{max}}{\dot\gamma}$}&
    \eqnum\label{eq:max_torque}\\[5pt]
    %\hline
    &Maximum rotational velocity&&\multicolumn{2}{c}{$\displaystyle\dot\gamma<\mathrm{F}_\mathrm{\gamma}\Omega_\mathrm{max}$}&
    \eqnum\label{eq:useful_space_rot}\\[5pt]
    %\hline
    &Secondary flow&&\multicolumn{2}{c}{$\displaystyle\eta>\frac{\rho\dot\gamma R^2}{4{\mathrm{F}_\mathrm{\gamma}}^3}$}&
    \eqnum\label{eq:sec_flow_viscosity}\\[5pt]
    \hline
    \multicolumn{2}{c|}{Gap-error formulation}&&\multicolumn{2}{c}{$\displaystyle\frac{h_\mathrm{c}}{\eta_\mathrm{m}} = \left(\frac{1}{\eta_\mathrm{r}}\right)h_\mathrm{c} + \frac{\epsilon}{\eta_\mathrm{r}}$}&
    \eqnum\label{eq:gap_error_estimate}\\[5pt]
    \hline
    \end{tabularx}}
\end{table}

\subsection{Considerations on suspensions and magnetorheological measurements}

With regard to suspensions, several phenomena must be taken into account. Here, we focus on dilute/semi-dilute suspensions of spherical particles (volume fractions below 0.25). Brownian motion, inertial effects, and the overall ability of the particles to follow the fluid flow are important and should be evaluated. The Peclet ($Pe$), (particle) Reynolds ($Re_\mathrm{p}$) and Stokes ($St$) numbers are used to quantify these respective effects which may be generally disregarded if the following adimensional quantities are sufficiently small\citep{mewis2012}:
\begin{equation}
    \frac{1}{Pe}=\frac{k_B\,T}{6\,\pi\,\eta\,(d_\mathrm{p}/2)^3\,\dot\gamma}\,,
    \label{eq:Peclet}
\end{equation}
\begin{equation}
    Re_\mathrm{p}=\frac{\rho\,(d_\mathrm{p}/2)^2\,\dot\gamma}{\eta}\,,
    \label{eq:Reynolds}
\end{equation}
\begin{equation}
    St=\frac{\rho_\mathrm{p}\,d_\mathrm{p}^2\,\dot\gamma}{18\,\eta}\,.
    \label{eq:Stokes}
\end{equation}
$d_\mathrm{p}$ and $\rho_\mathrm{p}$ are the particle diameter and density, respectively, $k_B$ is the Boltzmann constant and $T$ is the temperature. Additionally, gravitational effects can be relevant if the density difference between the dispersed and continuous phases is significant, and particle migration may occur from shear rate gradients within the flow or particle surface roughness and collisions\citep{mewis2012}. This particle movement can provoke concentration gradients, originating depletion layers near the solid boundaries, or even shear banding. Particle depletion near the geometry walls provokes an apparent slip of the dispersed phase and is common in suspension rheology, resulting in a viscosity gap-dependence. This effect may be corrected a posteriori or counteracted by increasing the surface roughness of the geometry or detailing it (milled or serrated, for example)\citep{buscall2010,yoshimura1988}. Suspensions may also display time-dependent behaviour as the microstructure takes time to adjust to the applied shear. Moreover, the rheological response also depends on the shear history, particularly for low shear rates or after a resting period where the suspension may achieve a metastable state\citep{mewis2012}. 

Concerning magnetorheological measurements, the application of an external magnetic field induces magnetic dipoles in the individual particles, causing them to aggregate into elongated chains aligned with the magnetic field lines\citep{mewis2012}. As such, a reversible and controllable microstructure is enabled through which the bulk rheological properties can be significantly altered\citep{lopez2006}. Additional experimental concerns arise from the time-dependent nature of the structure formation/destruction mechanisms and heating from magnetic field generation, which can be significant for strong fields.

\section{Materials and methods}
\label{sec:methods}

\subsection{Rheometer and accessories}
\label{sec:subsec:setup}

\begin{figure}[htp]
\captionsetup[subfigure]{labelformat=empty}
  \sbox0{\begin{minipage}[b]{0.3\linewidth}
        \centering
        (a) MCR302-e
        \subfloat[]{\includegraphics[trim={2cm 0cm 2cm 0cm},clip,width=\linewidth]{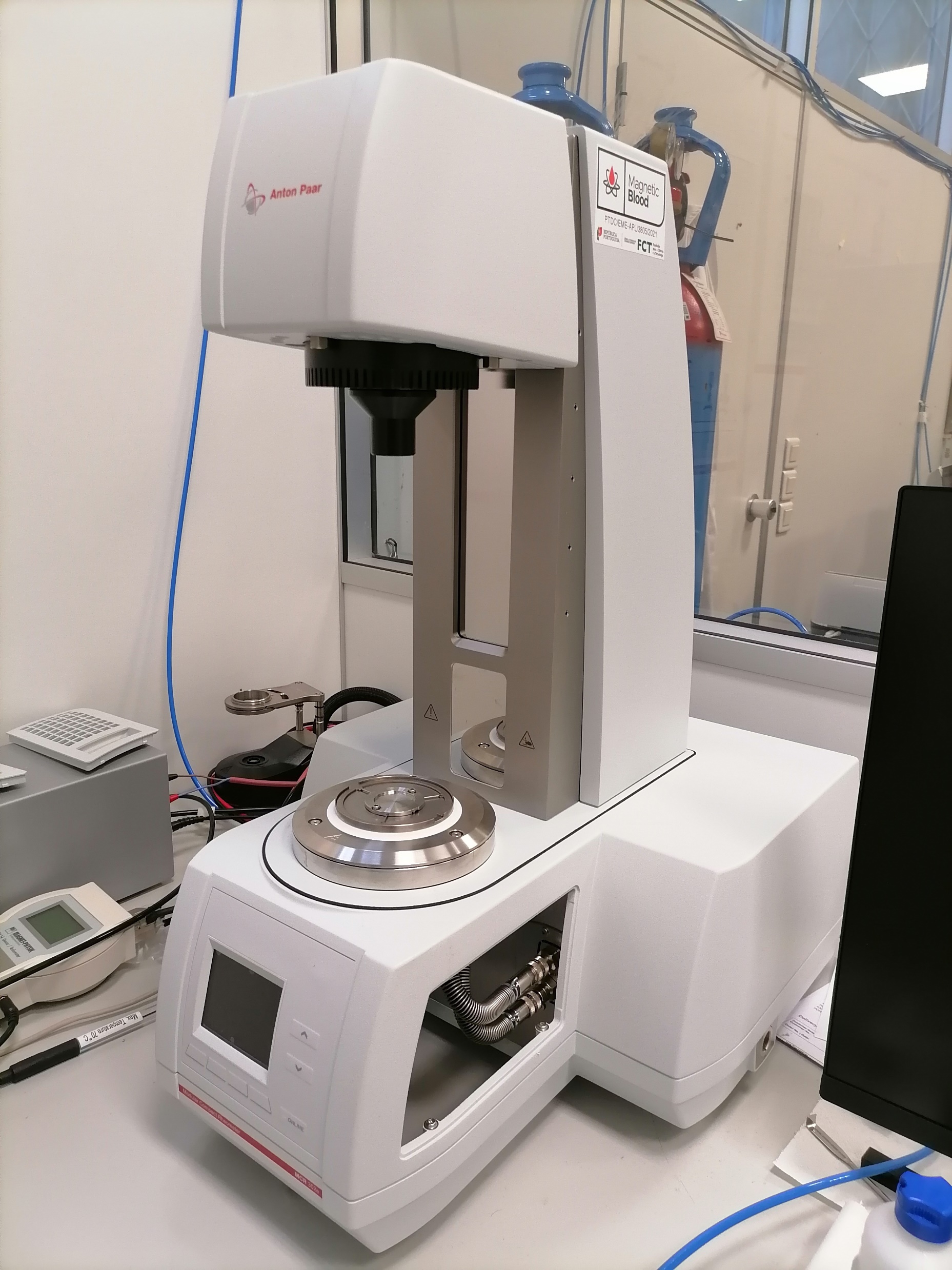}}
    \end{minipage}}
    \mbox{\usebox0}
    \hfill
    \mbox{\begin{minipage}[b][\ht0][s]{0.3\linewidth}
        \centering
        (b1) SP
        \subfloat[]{\includegraphics[trim={1cm 3cm 1cm 7cm},clip,width=\linewidth]{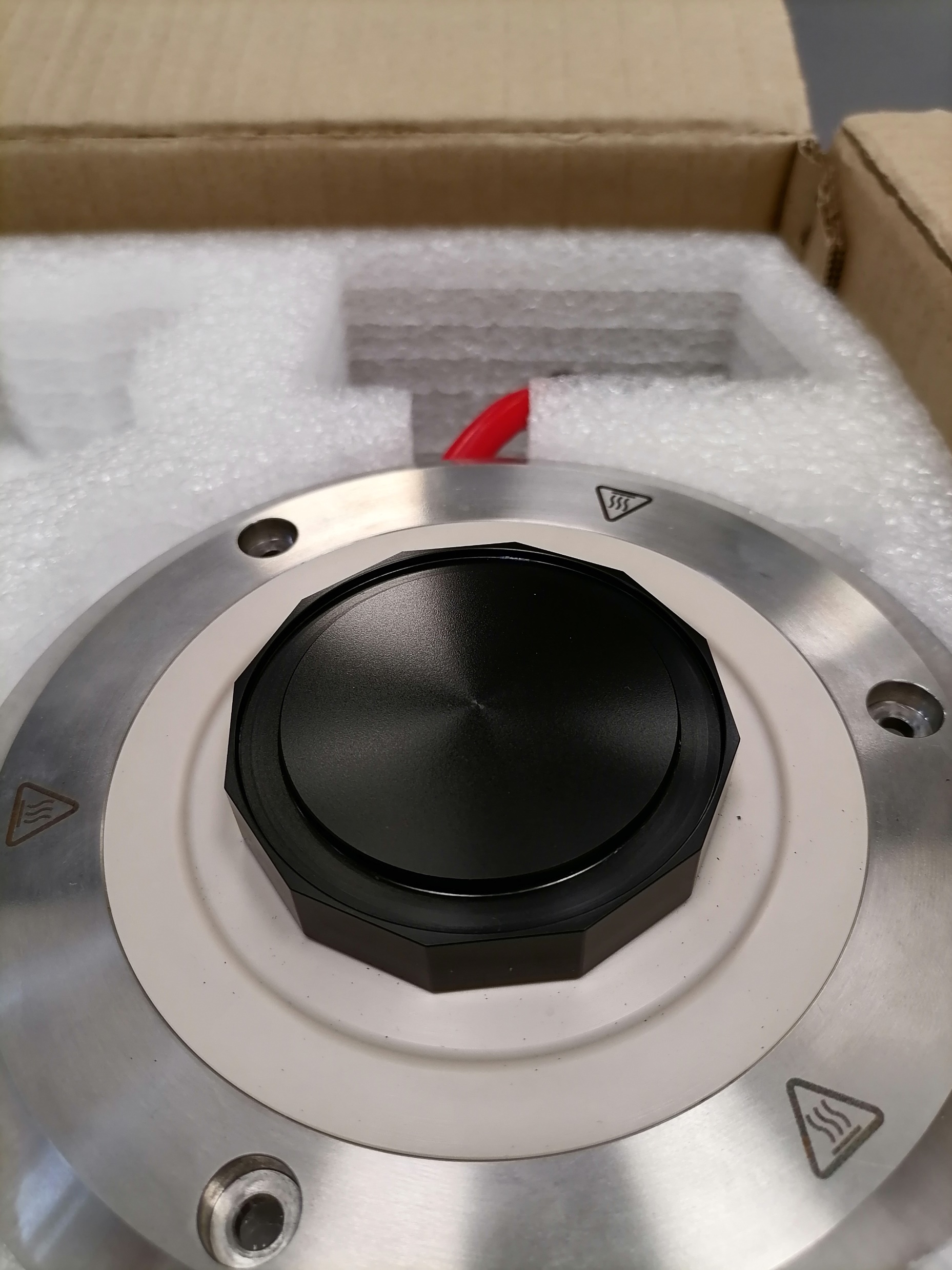}}
        \vfill
        \centering
        (b2) MP
        \subfloat[]{\includegraphics[trim={1cm 4cm 1cm 6cm},clip,width=\linewidth]{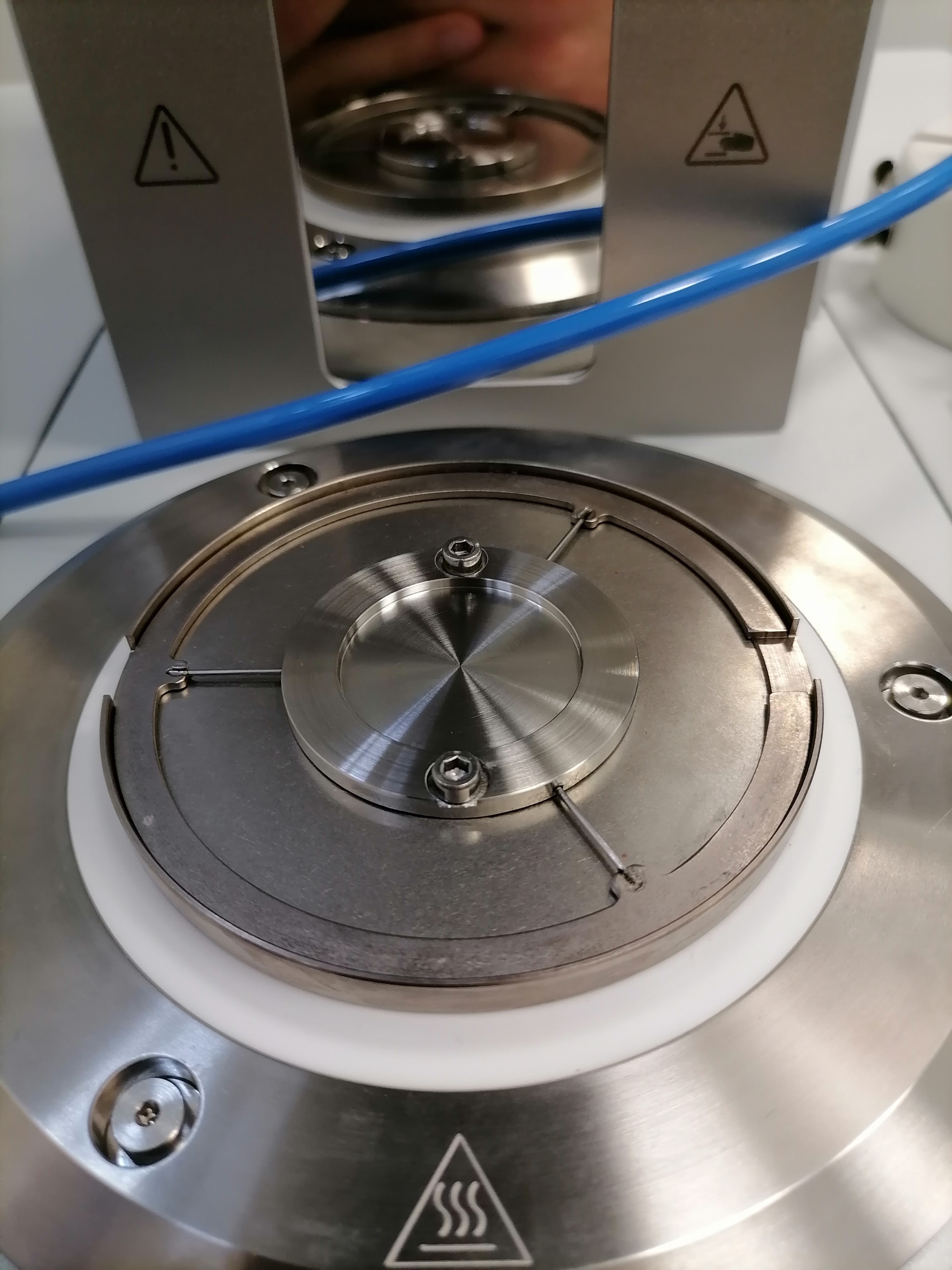}}
    \end{minipage}}
    \hfill
    \mbox{\begin{minipage}[b][\ht0][s]{0.3\linewidth}
        \centering
        (c1) PP20 MRD
        \subfloat[]{\includegraphics[trim={1cm 2cm 1cm 8cm},clip,width=\linewidth]{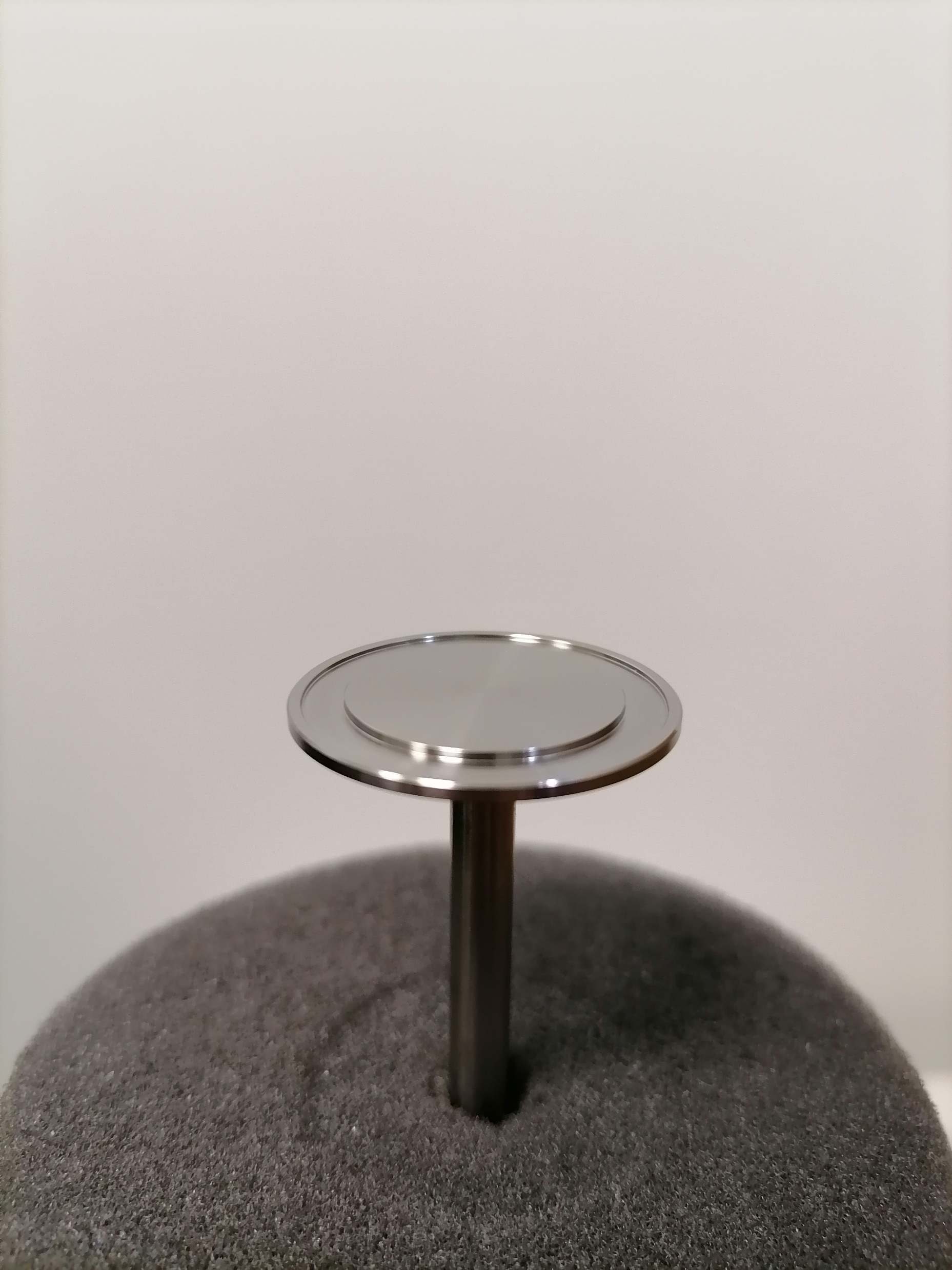}}
        \vfill
        \centering
        (c2) CP20 MRD
        \subfloat[]{\includegraphics[trim={1cm 0.5cm 1cm 9.5cm},clip,width=\linewidth]{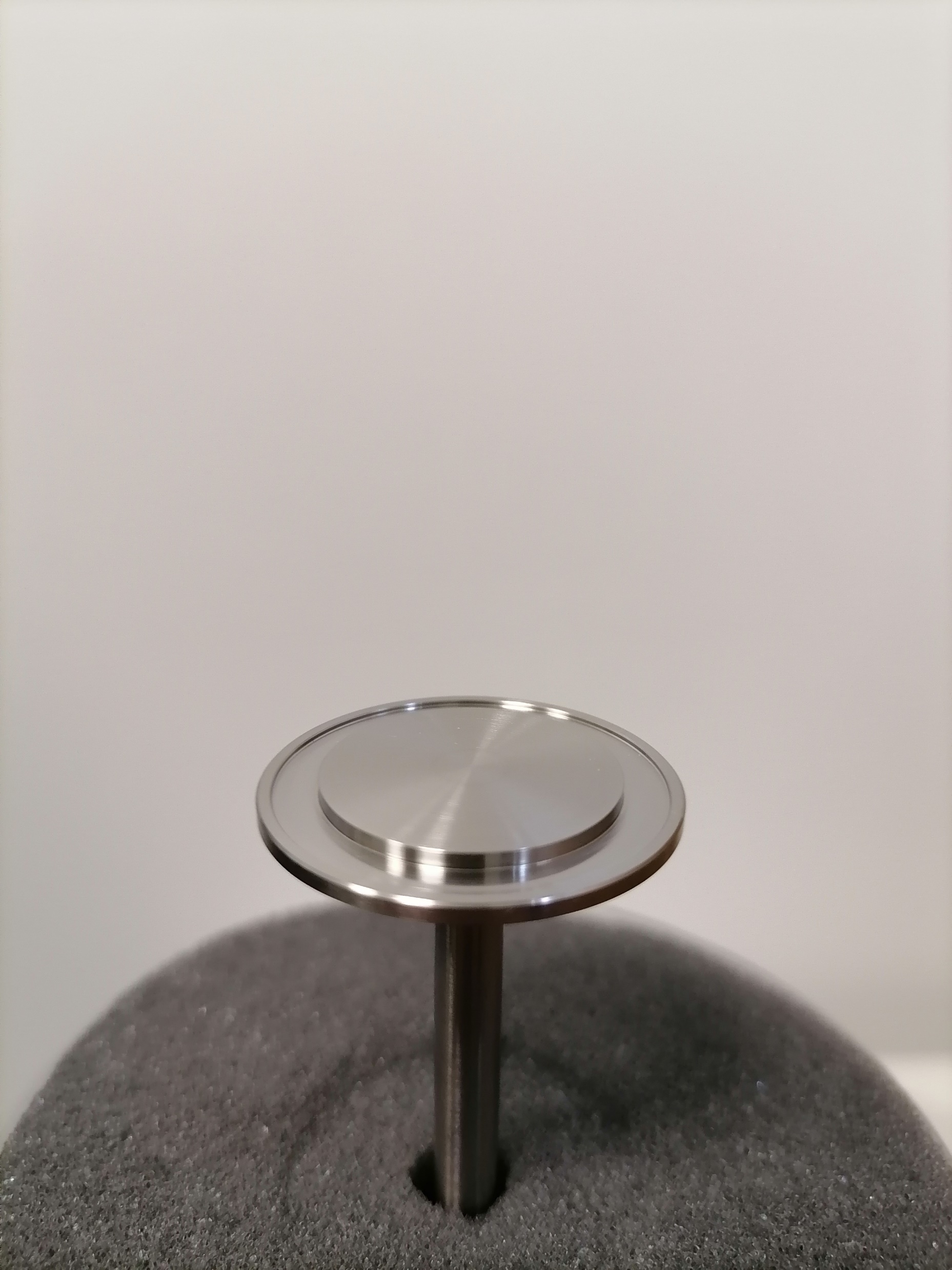}}
    \end{minipage}}
\caption{Experimental setup: (a) Anton Paar MCR302-e rotational rheometer, (b) SP and MP bottom plates, and (c) PP20 MRD and CP20 MRD geometries.}
\label{fig:Setup}
\end{figure}

An Anton Paar MCR302-e stress-controlled rotational rheometer was used for the rheological characterisation (Figure \ref{fig:Setup}(a)). According to the supplier, the instrument's limitations were: $\mathrm{T}_\mathrm{min}=1$ nN$\cdot$m, $\mathrm{T}_\mathrm{max}=230$ mN$\cdot$m and $\Omega_\mathrm{max}=314$ rad/s.  

Two static bottom plates were used; one was designed explicitly for magnetorheological measurements (Magnetic Plate, or MP) and the other for standard testing (Standard Plate, or SP), both are shown in Figure \ref{fig:Setup}(b). The SP had a diameter of 50 mm and allowed for careful temperature control through a Peltier system. On the other hand, the MP had a diameter of about 30 mm and was attached to a magnetorheological cell, which fit into the rheometer and was connected to a power source (being responsible for the generation of the magnetic field). An external cooling system was used to dissipate heat from the magnetic field generation, but keeping a precise temperature control within the fluid sample still remained challenging. The MP itself was fixed to the cell by two screws and centred by three radial fixtures, as can be seen in Figure \ref{fig:Setup}(b2). 

\begin{figure}[htp]
\centering
\includegraphics[width=\linewidth]{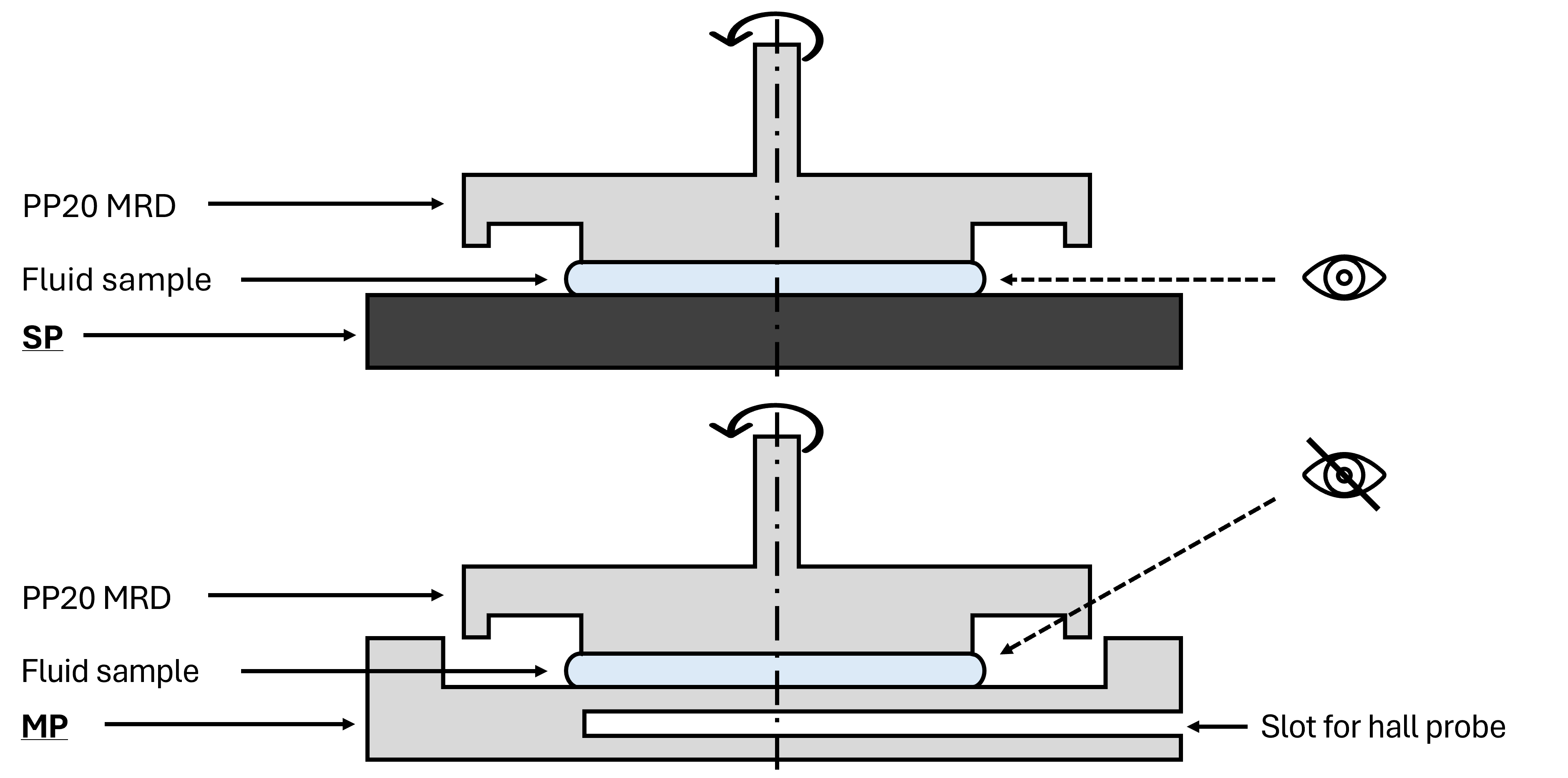}
\caption{Schematic of the PP20 MRD geometry with either bottom plate (not to scale) and associated difficulties with evaluating the loading quality.}
\label{fig:Setup - schematic}
\end{figure}

Two measuring geometries were employed: one parallel-plates (PP20 MRD) and one truncated cone-plate (CP20 MRD), specifically designed for magnetorheological measurements. Both had a diameter of 20 mm, and the CP had a cone angle of 1.981$^\circ$ and a truncation height of 0.084 mm. Both geometries are shown in Figure \ref{fig:Setup}(c). Either MRD geometry had an outer region consisting of two rings with different depths (both lesser than the plate itself) that hindered the visualisation of the loading quality. In the SP, we could still visualise the sample's profile with a high-contrast background (white walls in our case). On the other hand, the MP also had an outer rim, which completely blocked our vision of the contact line in the MRD geometries, impeding the evaluation of the loading quality. Figure \ref{fig:Setup - schematic} shows a schematic of the PP20 MRD with either bottom plate.

\subsection{Working fluids}

Four different Newtonian fluids were used: a calibration oil provided by Anton Paar (Cal. Oil), two aqueous solutions of glycerol in mass concentrations of 75.20 and 93.05 wt\% (75.20Gly and 93.05Gly, respectively), and an aqueous solution of 52 wt\% of Dimethyl sulfoxide (DMSO) (Newtonian blood analogue\citep{campo2013}, NBa). Table \ref{tab:fluids} gives each working fluid's composition, expected density and viscosity at the experimental temperature of 20$^\circ$C.

\begin{table}[htp]
    \centering
    \renewcommand{\arraystretch}{1.5}
    \small{
    \caption{Composition and expected viscosity, $\eta$, and density, $\rho$, of the working fluids (at 20$^\circ$C). Calibration oil data from the supplier (Anton Paar). Glycerol and water estimates from \citet{volk2018}. DMSO data from \citet{budeanu2021}}
    \label{tab:fluids}
    \begin{tabularx}{\textwidth}{Y|YYYY|YY}
    \hline
    Sample& 
    Water [wt\%]&
    Glycerol [wt\%]&
    Cal. Oil [wt\%]&
    DMSO [wt\%]&
    $\rho$ [kg/m\textsuperscript{3}] &
    $\eta$ [mPa$\cdot$s]\\
    \hline
    \hline
    Cal. Oil&
    -&
    -&
    100&
    -&
    816.1&
    3.66\\
    75.20Gly&
    24.80&
    75.20&
    -&
    -&
    1194.9&
    36.98\\
    93.05Gly&
    6.95&
    93.05&
    -&
    -&
    1243.0&
    365.68\\
    NBa&
    48.00&
    -&
    -&
    52.00&
    1051&
    3.32\\
    \hline
    \end{tabularx}}
\end{table}

\subsection{Experimental procedure}

\subsubsection{Pre-measurement procedure}
Before any measurement, the zero-gap was set, the inertia of both the drive and the measuring system was acquired, and a motor adjustment function was conducted according to the RheoCompass\texttrademark{} (Anton Paar) procedure. The set zero-gap and inertia call operations were repeated at every rheometer reset and bottom plate/geometry switch. Both plates were thoroughly cleaned of any previous sample residues and dust particles, cleansed with ethanol and dried with compressed air. The sample volume (previously determined and given in Table \ref{tab:volumes}) was accurately measured with a VWR\textsuperscript{\textregistered{}} standard line precision pipette and carefully applied at the centre of the bottom plate. The calibration oil's low surface tension required the sample to be placed on the geometry instead, or a combination of both. The rheometer head was then lowered with the predefined viscoelastic movement profile. When possible, the loading quality was evaluated for any signs of under/overfilling and contact line asymmetry.

\subsubsection{Data acquisition and processing}
Viscosity curves were obtained by imposing a shear rate logarithmic ramp-up with about 10 points per decade, between 10\textsuperscript{-1} and 10\textsuperscript{4} s\textsuperscript{-1}. A constant waiting time of a few seconds was defined between shear rate shift and data acquisition, guaranteeing steady flow. For all measurements, at least 5 tests were conducted for statistical robustness, posteriorly averaged, and 95\% confidence intervals were calculated using Student's t-distribution.

\subsubsection{Temperature control}
With the SP, the Peltier system was set to maintain a constant temperature (20$^\circ$C), whereas no such fine temperature control was achievable with the MP. Our approach was to conduct the measurements when the room temperature did not diverge too much from the analogous tests with the SP (maximum $\pm\;5^\circ$C) while keeping the average temperature of the measurement set close to the targeted 20$^\circ$C.

\subsection{Magnetorheological measurements}
\label{subsec:Methods_magnetorheological}

The magnetorheological cell was able to generate a magnetic field perpendicular to the flow direction, which was measured with an FH 54 teslameter (MAGNET-PHYSIK) inserted into a slot underneath the MP (depicted in Figure \ref{fig:Setup - schematic}). The geometries (PP20 MRD and CP20 MRD) were non-magnetic, which prevented radial forces from acting on the shafts, and the magnetic circuit was closed with a yoke placed on top of the geometry \citep{laeuger2005}.

The measurements were conducted with the NBa working fluid seeded with M-270 Carboxylic Acid Dynabeads\texttrademark\,(Thermo Fisher Scientific) paramagnetic particles. These had a diameter of $d_\mathrm{p}=2.8$ \textmu m, density of $\rho_\mathrm{p}=1600$ kg/m\textsuperscript{3} and their magnetisation curve is given by \citet{grob2018} (saturation of $M_\mathrm{sat} = 6.4$ Am\textsuperscript{2}/kg for $B \gtrsim 500$ mT). Three mass concentrations were tested: 5, 10 and 15 wt\%, which correspond to dilute/semi-dilute suspensions\citep{mewis2012} with volume fractions of $\phi\approx3.4, 6.9$ and $10.6$ vol\%.

Prior to loading, the samples were thoroughly mixed to redisperse the sedimented particles and to break any possible microstructure that may have formed at rest. The steady shear viscosity data was gathered at a constant shear rate of $\dot\gamma=500$ s\textsuperscript{-1} and 10 magnetic field density values were applied up to $B\leq720$ mT. The time interval between field density change and data acquisition was set to 20 s to avoid gathering data before the magnetic-induced microstructure stabilised.

\section{Results and discussion}
\label{sec:results}

The nomenclature used to refer to the bottom plates and geometries is surmised in Table \ref{tab:nomenclature}. The used bottom plates (SP and MP) and geometries (PP20 MRD and CP20 MRD) are all shown in Figure \ref{fig:Setup} and were discussed in Subsection \ref{sec:subsec:setup}.
\begin{table}[htp]
    \centering
    \renewcommand{\arraystretch}{1.5}
    \small{
    \caption{Nomenclature used to refer to the employed bottom plates and geometries}
    \label{tab:nomenclature}
    \begin{tabularx}{\textwidth}{Y|Y}
    \hline
    Bottom plates& 
    Geometries\\
    \hline
    \hline
    SP: Standard Plate&
    PP: parallel-plates\\
    MP: Magnetic Plate&
    CP: cone-plate\\
    \hline
    \end{tabularx}}
\end{table}

\subsection{Standard measurements}

Previous experimental results pointed towards a non-negligible gap-error of the MP. Larger sample volumes are required to fill geometries on the MP than on the SP (these are given in Table \ref{tab:volumes}), and a viscosity dependence on the gap height was found, with lower measured viscosities for smaller gaps. The discussion on these preliminary measurements is presented as supplementary material. 

\begin{table}[htp]
    \centering
    \renewcommand{\arraystretch}{1.5}
    \small{
    \caption{Employed sample volumes for either geometry (PP20 MRD at different gap heights, $h_\mathrm{c}$, and CP20 MRD) in either bottom plate (SP and MP)}
    \label{tab:volumes}
    \begin{tabularx}{\textwidth}{YY|YYYYYYY|YY}
    \hline
    \multicolumn{11}{c}{Sample volume [\textmu l]}\\
    \hline
    \multicolumn{2}{c|}{\multirow{2}{*}{\centering Bottom plate}}&
    \multicolumn{7}{c|}{PP20 MRD (for $h_\mathrm{c}$ in mm)}&
    \multicolumn{2}{c}{\multirow{2}{*}{\centering CP20 MRD}}\\
    &&
    0.05&
    0.10&
    0.15&
    0.20&
    0.25&
    0.30&
    0.35&
    &
    \\
    \hline
    \hline
    \multicolumn{2}{c|}{SP}&
    20&
    36&
    51&
    67&
    82&
    98&
    114&
    \multicolumn{2}{c}{76}\\
    \multicolumn{2}{c|}{MP}&
    27&
    42&
    58&
    73&
    89&
    105&
    120&
    \multicolumn{2}{c}{84}\\
    \hline
    \end{tabularx}}
\end{table}

\begin{figure}[htp]
\centering
\includegraphics[width=\linewidth]{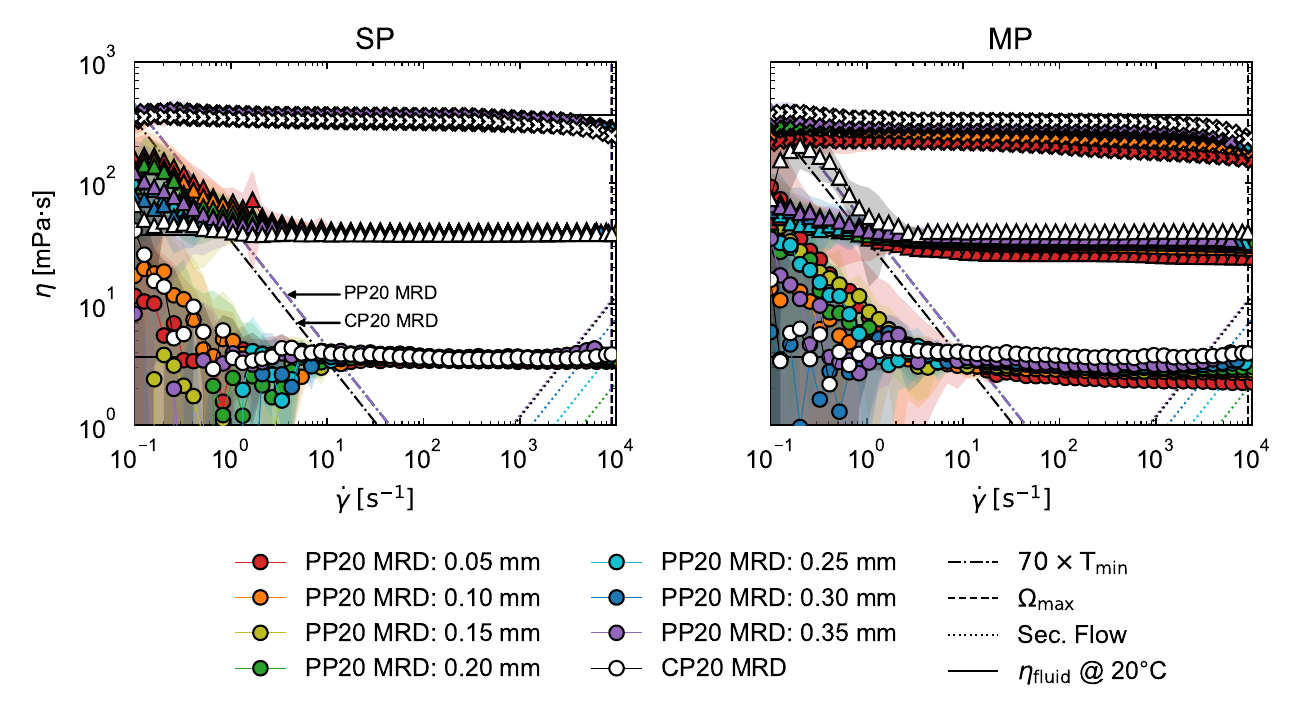}
\caption{Viscosity curves obtained with three Newtonian fluids: Cal. Oil (circular markers), 75.20Gly (triangular markers) and 93.05Gly (cross markers), on either bottom plate: (left) SP and (right) MP. Data gathered with the CP20 MRD and PP20 MRD (gap heights between 0.05 and 0.35 mm). With the MP, the PP-measured viscosity curves show a clear gap-dependence, with lesser gaps yielding lower viscosities. On the other hand, the CP returns slightly larger viscosities on the MP than on the SP.}
\label{fig:Flow curves}
\end{figure}

We set out to test the measurement quality with three Newtonian fluids of different viscosities (Cal. Oil, 75.20Gly and 93.05Gly) on both geometries (PP20 MRD and CP MRD) and both bottom plates (SP and MP). The PP20 MRD gap height was varied between 0.05 and 0.35 mm and Figure \ref{fig:Flow curves} shows the obtained viscosity curves. The surrounding error bands represent the 95\% confidence intervals. The measured viscosity was plotted along with the experimental limits associated with the rheometer's specifications: torque, $\mathrm{T}_\mathrm{max/min}$, and rotational velocity, $\Omega_\mathrm{max}$, and the prediction of secondary flow onset (expressions \ref{eq:min_torque}-\ref{eq:sec_flow_viscosity}); if any are not visible it is because they fall out of the experimental window and are, thus, not relevant for those particular measurements. The secondary flow limit is dependent on the geometry configuration (see Equation \ref{eq:sec_flow_viscosity}); as such, a secondary flow limit has been plotted for each geometry and tested PP20 MRD gap, using the same colour scheme that describes the corresponding experimental data.

The results obtained with the SP show generally constant viscosities for all three tested fluids, with no significant differences between either geometry or PP gap heights. Large errors can be seen at low shear with the two least-viscous fluids, which should be associated with low-torque issues. Because the rheometer's minimum torque limit is outside the experimental window (at lower shear rates for the shown viscosity range), we believe these low-shear errors are probably due to surface tension torque from contact line asymmetry\citep{johnston2013}. Applying an adjustment factor of 70 to the minimum-torque limit ($70\times\mathrm{T}_\mathrm{min}$) seems to predict the low-shear errors accurately, but significantly narrows the experimental window. Despite our efforts to carefully evaluate the correct sample volumes, the design of the MRD geometries makes this task challenging, as it hinders the visualisation of the contact line during measurement. As a result, small loading errors may have gone unnoticed. Nevertheless, the adjusted torque is, despite significant, not shocking ($70\times\mathrm{T}_\mathrm{min}=70$ nN$\cdot$m). \citet{johnston2013} reported measured torques with water up to 1 \textmu N$\cdot$m at low shear from loading errors. Other works, despite employing lesser adjustment factors also had less torque resolution at low shear\citep{soulages2009,oliveira2006,rodd2005}.

At high shear, the Cal. Oil presents a slight viscosity increase on the larger PP gaps and the CP, which is related to secondary flow onset and is accurately predicted by the respective limit\citep{turian1972}. The most viscous solution shows evidence of viscous heating, visible through an acute viscosity decrease at high shear (there was no evidence of sample loss). Even though the SP can guarantee a near-constant temperature via the Peltier system, the temperature sensor is not in direct contact with the fluid sample. Thus, delays in temperature control are not surprising when dealing with very sharp temperature variations in the sample. Additionally, with the 93.05Gly solution a slight viscosity decrease is noted throughout the whole viscosity curve. At low to medium shear this should not be a symptom of viscous heating, but it might be due to water absorption over time, which may be more noticeable for small gaps\citep{ault2023}. The same effect may not be as noticeable for the 75.20Gly due to its lesser glycerol concentration.

With the MP, the adjusted low-torque and secondary flow limits also predict the low-shear uncertainty and high-shear measured viscosity increment. Moreover, the viscous heating effects on the 93.05Gly are not significantly more pronounced. Compared to the SP data, on the MP, the CP20 MRD returned slightly larger viscosities, while the PP20 MRD displays a clear viscosity reduction with gap decrease.

\begin{figure}[htp]
\centering
\includegraphics[width=\linewidth]{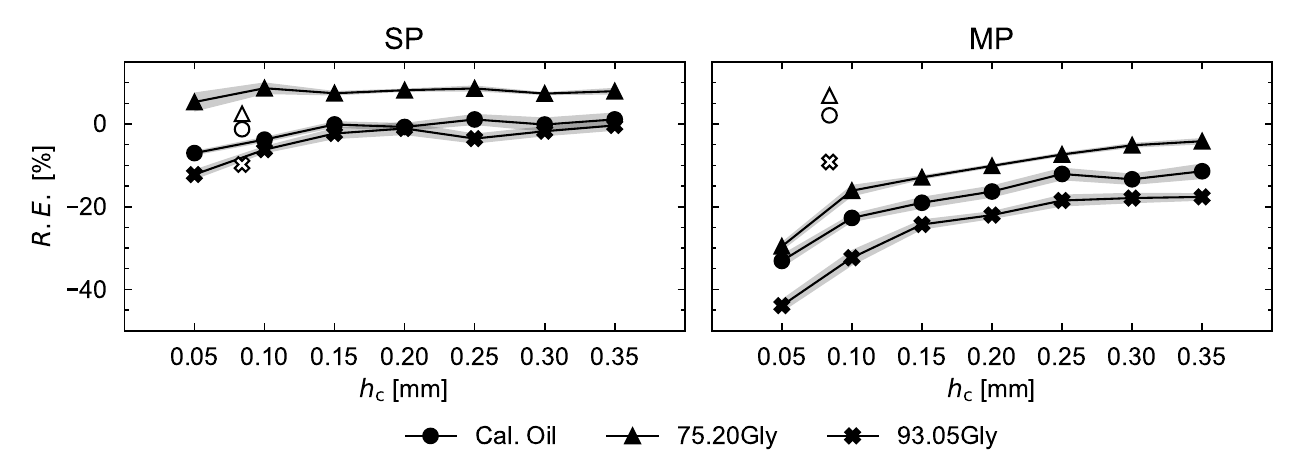}
\caption{Mean viscosity relative errors and respective 95\% confidence intervals for all three fluids: Cal. Oil (circular markers), 75.20Gly (triangular markers) and 93.05Gly (cross markers), gathered on both geometries: PP20 MRD (filled markers) and CP20 MRD (empty markers), and on either bottom plate: (left) SP and (right) MP. Confidence intervals for the CP20 MRD always within  marker size. On the MP, there is a clear gap-dependence of the PP-measured viscosity and a slight CP overestimation.}
\label{fig:Mean_relative_errors}
\end{figure}

Focusing on the PP20 MRD results, we can conduct a more in-depth analysis by evaluating useful data at multiple gap-heights. To avoid experimental errors, the data selected from each fluid was limited to a particular shear rate range that avoided low-shear uncertainties and secondary flow/viscous heating issues at high shear: $20\leq\dot\gamma_\mathrm{Cal. Oil}\leq2000$, $3\leq\dot\gamma_\mathrm{75.20Gly}\leq6000$ and $0.2\leq\dot\gamma_\mathrm{93.05Gly}\leq3000$ s\textsuperscript{-1}. The data obtained at each gap was averaged over the selected shear rate range, $\overline{\eta_\mathrm{m}}$, and compared to the expected value, $\eta_\mathrm{exp.}$ (Table \ref{tab:fluids}), through a relative error:
\begin{equation}
    \mathrm{R.E.} = \frac{\overline{\eta_\mathrm{m}}-\eta_\mathrm{exp.}}{\eta_\mathrm{exp.}}\,,
\end{equation}
which is shown, for each fluid, in Figure \ref{fig:Mean_relative_errors}.

On the MP, the measured viscosity dependence with PP gap height is evident. For the lowest gap of 0.05 mm, the relative error is significant, between -45 and -30\%, but it diminishes with increasing gap height to errors between -18 and -4\% for $h_\mathrm{c} =0.35$ mm. With the CP20 MRD, however, positive relative errors are noticed for the two least viscous fluids (2 and 7\%), while the most viscous solution (93.05Gly) returned a negative error of -9\%. It is important to remember that with the 93.05Gly solution, we did notice a viscosity decrease along the curve, possibly due to water absorption, which may be why this fluid shows more significant relative errors. 

Although the SP's relative errors are not as striking as the MP's, we can still notice a slight dependence of the PP-measured viscosity with gap height. The Cal. Oil and 93.05Gly tend to a null relative error with gap increase, while the 75.20Gly solution presents a weak tendency towards a 7\% positive error which could be due to a inaccurate preparation of the solution, where the glycerol concentration was slightly larger than intended. The CP20 MRD returned lesser errors on the SP than on the MP, with the two less viscous fluids approaching a null error.

\begin{figure}[htp]
\centering
\includegraphics[trim={0cm 0cm 0cm 2cm},clip,width=\linewidth]{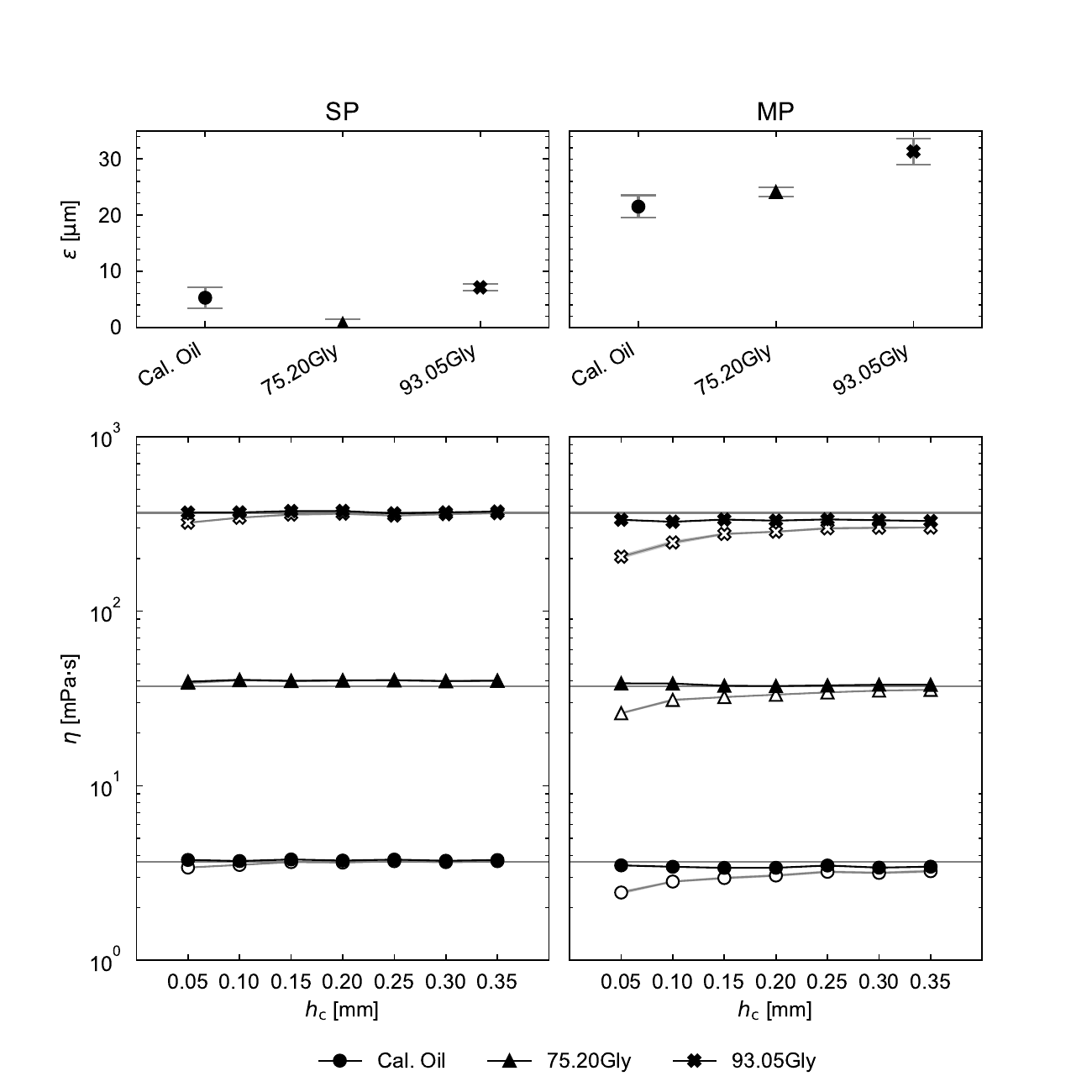}
\caption{Gap-error, $\varepsilon$, and true viscosity estimates, $\eta_\mathrm{r}$ (filled markers, with averaged measured viscosity, $\overline{\eta_\mathrm{m}}$, in open markers for comparison) for all three fluids: Cal. Oil (circular markers), 75.20Gly (triangular markers) and 93.05Gly (cross markers), on either bottom plate: (left) SP and (right) MP. The MP has an associated gap-error around $\varepsilon_\mathrm{MP}\approx22.8$ \textmu m.}
\label{fig:Gap-error_correction}
\end{figure}

Because this viscosity gap-dependence is not observed on the SP, we believe there is a systematic gap-error on the MP and that slip effects are negligible. As such, fitting this gap-varying data to Equation \ref{eq:gap_error_estimate}, we can estimate the gap-error, $\varepsilon$, and correct the measured viscosity to an estimated value\citep{kramer1987}: $\eta_\mathrm{r} = \overline{\eta_\mathrm{m}}\;(1+\varepsilon/h_\mathrm{c})$. The fitting procedure was done using Python's Statsmodels GLM functionalities, and Figure \ref{fig:Gap-error_correction} shows the estimated gap-errors and corrected viscosities for each fluid.

On the SP, the 75.20Gly solution returned a practically null gap-error, while small values are observed for the Cal. Oil and the 93.05Gly. These non-zero estimated gap-errors may have arisen from measurement uncertainty with the low-viscosity Cal. Oil, whereas for the high-concentration glycerol solution it may be a result of the previously-mentioned water absorption. From these results we assume hereafter that there is no gap-error on the SP. Contrarily, on the MP the Cal. Oil and 75.20Gly point towards a gap error of approximately 22.8 \textmu m, while the 93.05Gly solution again presents a larger estimation. The corrections on the MP returned slightly lesser viscosities than on the SP, which could be due to deviations of the mean room temperature from 20$^\circ$C during measurement (the recorded mean temperatures were 21.1, 20.3 and 20.8$^\circ$C, for the Cal. Oil, 75.20Gly and 93.05Gly), but, overall, the estimates seem to accurately predict the fluids' true viscosities.

\begin{figure}[htp]
\centering
\includegraphics[width=\linewidth]{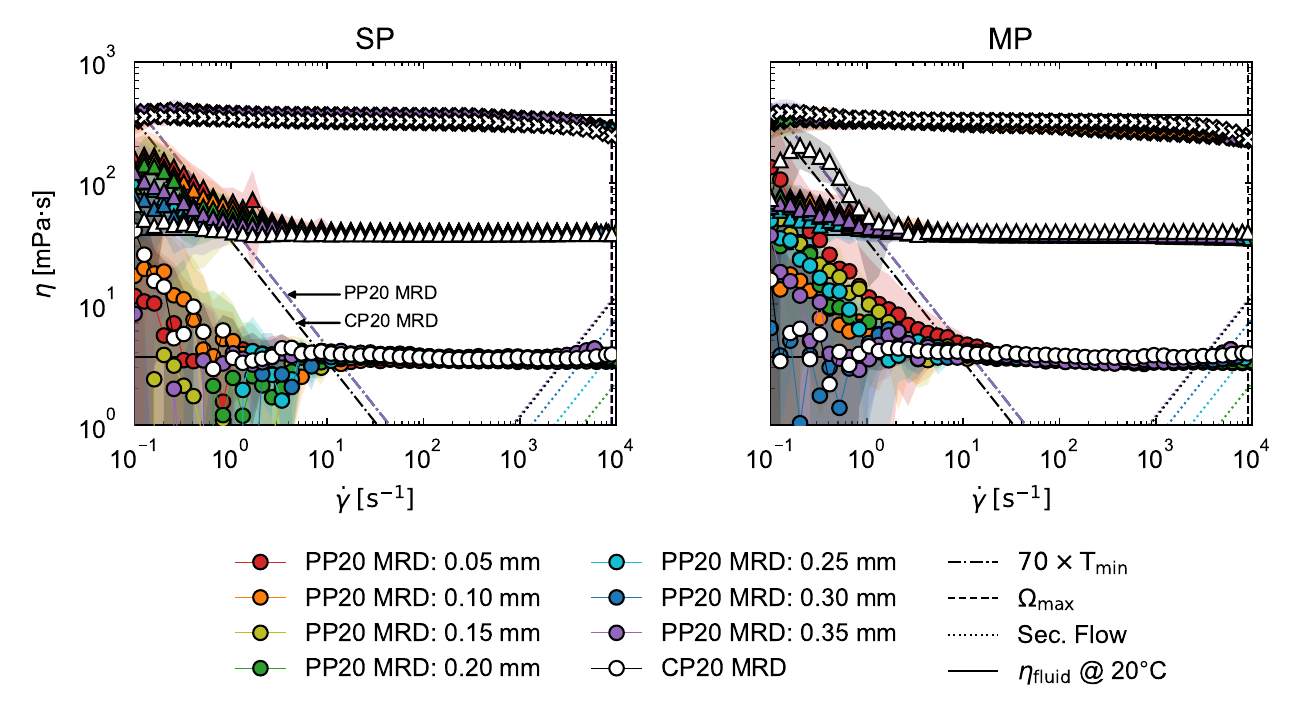}
\caption{Viscosity curves corrected for the gap error ($\varepsilon_\mathrm{SP}=0$ and $\varepsilon_\mathrm{MP}\approx22.8$ \textmu m) obtained with three Newtonian fluids: Cal. Oil (circular markers), 75.20Gly (triangular markers) and 93.05Gly (cross markers), on either bottom plate: (left) SP and (right) MP. Data gathered with the CP20 MRD and PP20 MRD (gap heights between 0.05 and 0.35 mm). The gap-error formulation is capable of correcting the PP data measured on MP.}
\label{fig:Flow curves - corrected}
\end{figure}

Once an efficient method to estimate the fluids' actual viscosities is validated, we can correct the experimental viscosity curves through two approaches. The first consists of correcting the viscosity by performing gap-error and true viscosity estimates at all shear rates, which could return some statistically interesting data, but it would also be accounting for errors associated with the experimental limits at extreme shear rates (either too low, from surface tension torque, or too high, from secondary flow onset and viscous heating). The alternative is to employ the previously obtained MP gap-error ($\varepsilon_\mathrm{MP}\approx22.8$ \textmu m) to correct the measured viscosity at all shear rates ($\eta_\mathrm{m}\;(1+\varepsilon/h_\mathrm{c})$). As the gap-error was estimated through viscosity data within the experimental window, we opted for this second method and Figure \ref{fig:Flow curves - corrected} shows the corrected viscosity curves (the SP data remained unaltered). The confidence intervals of the corrected curves were computed using the standard error of the experimental data, centred around the estimated true viscosities. As a result, the confidence intervals, plotted as error bands, reflect the experimental uncertainty, independent of the gap-error formulation.

Comparing to the raw viscosity curves (Figure \ref{fig:Flow curves}), the MP results are drastically altered, with the correction rectifying the experimental data effectively. Thus, hereafter we will focus on elucidating the cause of the gap-error and why the CP20 MRD seems immune to the measured viscosity reduction and prone to yielding larger values. 

Because the SP does not show such severe gap-error symptoms, issues regarding the setting zero-gap procedure seem unlikely, as it was the same for either bottom plate. The same can be said about surface finishing issues; even if the material for either plate is different, both were acquired from the same supplier (Anton Paar) and, thus, should have somewhat similar surface characteristics (the same reasoning for slip negligibility). Assuming that the MP is straight, as we hypothesise the SP is, the most simple cause should be an inclination of the MP that leads to geometry non-parallelism. 

\subsection{Numerical analysis of non-parallelism on standard measurements}
\begin{figure}[htp]
\centering
\begin{minipage}[t]{0.33\textwidth}
\centering
(a) PP20
\end{minipage}
\begin{minipage}[t]{0.61\textwidth}
\centering
(b) CP20
\end{minipage}
\includegraphics[width=\linewidth]{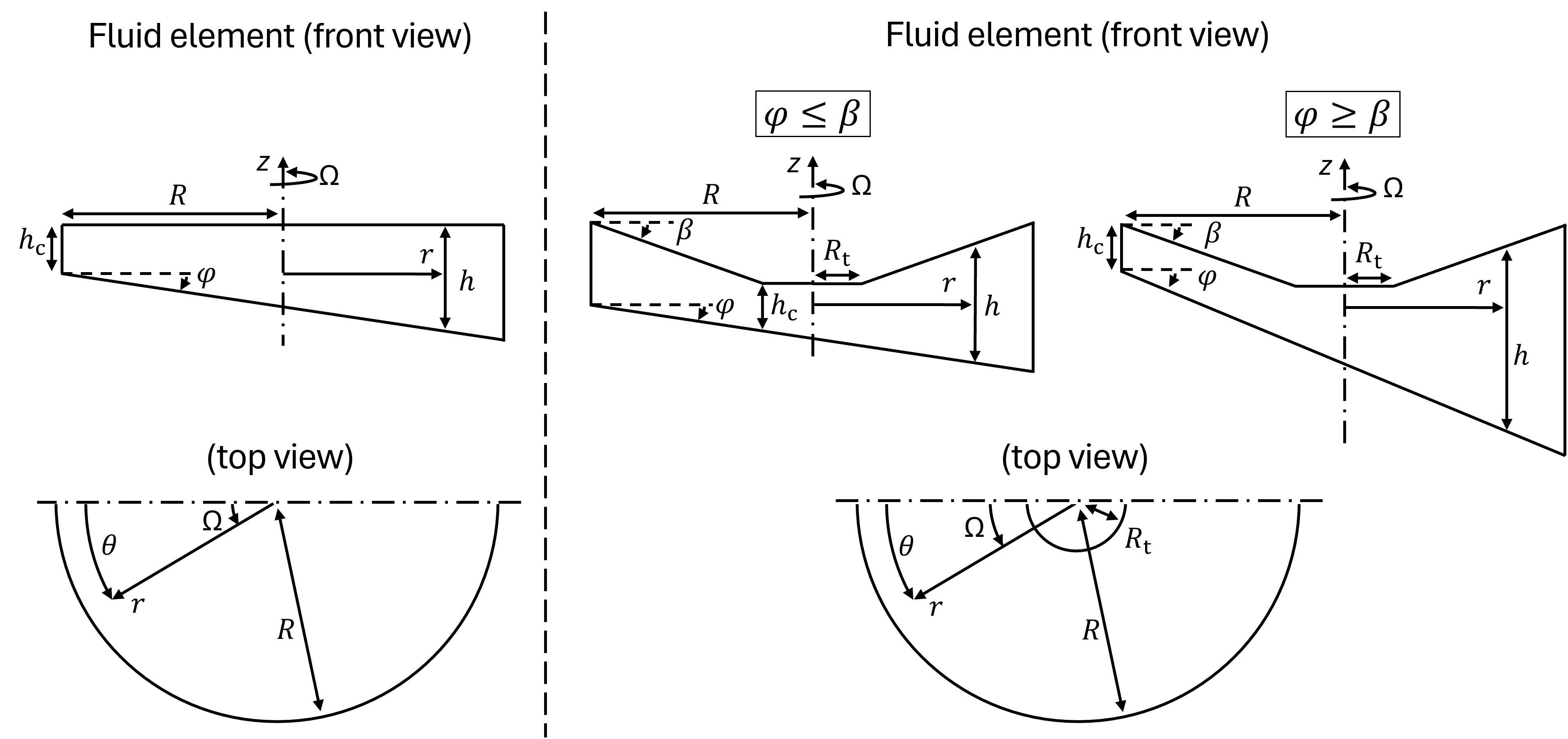}
\caption{Schematics (not to scale) of the modelled geometries with inclined bottom plate: (a) PP and (b) CP. The CP model differs depending on the magnitude of the inclination angle, $\varphi$, relative to the cone angle, $\beta$. Cylindrical coordinate system is assumed.}
\label{fig:Models}
\end{figure}

To validate our hypothesis, we set out to model both geometries (PP20 and CP20) with a varying inclination of the bottom plate to study them numerically in a 3D simulation. The model of a non-parallel PP is very straightforward, but a CP requires some consideration. When executing the zero-gap procedure, the rheometer assumes a zero gap when contact is made between the cone and the bottom plate. In our case, we have a truncated cone with an angle of $\beta=1.981^\circ$ and truncation height of 0.084 mm. If the inclination angle is lesser than the cone angle, $\varphi<\beta$, contact is made at the truncation radius, $\mathrm{R}_\mathrm{t}$. On the other hand, if $\varphi>\beta$, contact is made on the outer radius of the cone, $\mathrm{R}$, just like a PP geometry. When the inclination is equal to the cone angle, $\varphi=\beta$, both models are valid as contact is made along a radial segment. Figure \ref{fig:Models} shows schematics of the modelled PP and CP geometries (not to scale). $h_\mathrm{c}$ is the commanded gap, or the truncation height in the case of the CP, and the radius for both geometries was $R=20$ mm (PP20 and CP20).

The fluid flow was then calculated in COMSOL MultiPhysics, solving for the laminar, incompressible and isothermal flow of a Newtonian fluid of equal characteristics to the Cal. Oil at 20$^\circ$C (Table \ref{tab:fluids}). The steady state equations for conservation of mass and momentum are solved:
\begin{equation}
\rho\left(\mathbf{u}\cdot\mathbf{\nabla}\right)\mathbf{u}=-\mathbf{\nabla}p+\eta\mathbf{\nabla}^2\mathbf{u}+\mathbf{F}\,,
\end{equation}
\begin{equation}
    \mathbf{\nabla}\cdot\mathbf{u}=0\,.
\end{equation}
The applied boundary conditions were no-slip at the top and bottom walls. A rotational velocity, $\Omega$, was imposed at the top wall (the measuring system) while the bottom wall remained fixed. The sample/air interface was simplified in two ways. First, the shape of the interface was considered straight and vertical (Figure \ref{fig:Models}), while in real applications, we aim to have a perfect convex meniscus shape, whose contact angle with the solid parts is dependent on fluid surface tension. Second, the fluid/air interface was modelled as a free-slip condition for the sake of simplicity.

\begin{figure}[ht]
\captionsetup[subfigure]{labelformat=empty}
  \sbox0{\begin{minipage}[b]{0.3\linewidth}% measure height
        \centering
        (a) Top View
        \subfloat[]{\includegraphics[trim={5.2cm 5.2cm 0cm 0cm},clip,width=\linewidth]{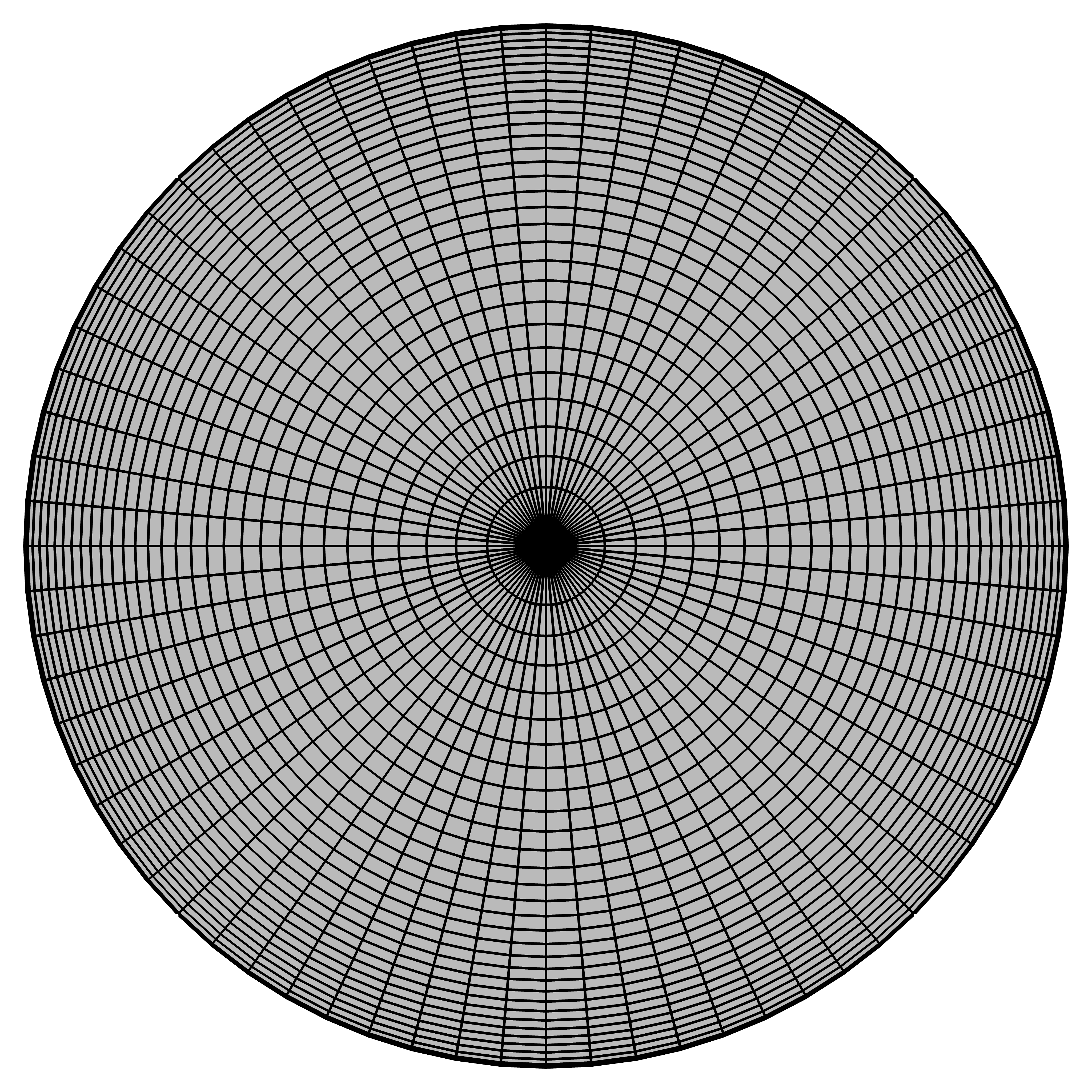}}
    \end{minipage}}%
    \mbox{\usebox0}%
    \hfill
    \mbox{\begin{minipage}[b][\ht0][s]{0.3\linewidth}
        \centering
        (b) Side view
        \subfloat[]{\includegraphics[trim={0cm 2cm 3cm 1cm},clip,width=\linewidth]{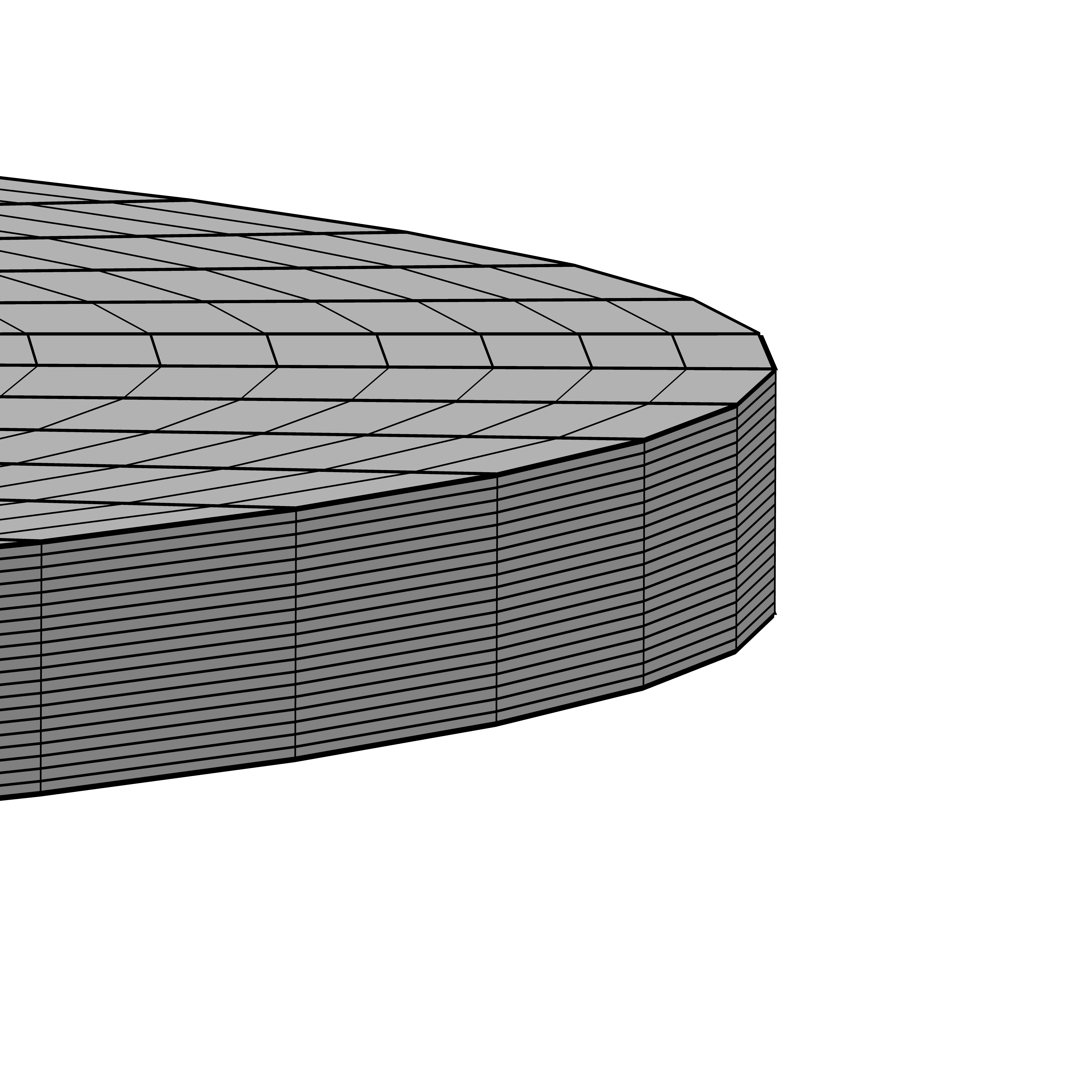}}
        \vfill
    \end{minipage}}
    \hfill
    \mbox{\begin{minipage}[b][\ht0][s]{0.3\linewidth}
        \centering
        (c) Inner region
        \subfloat[]{\includegraphics[trim={2cm 2cm 2cm 2cm},clip,width=\linewidth]{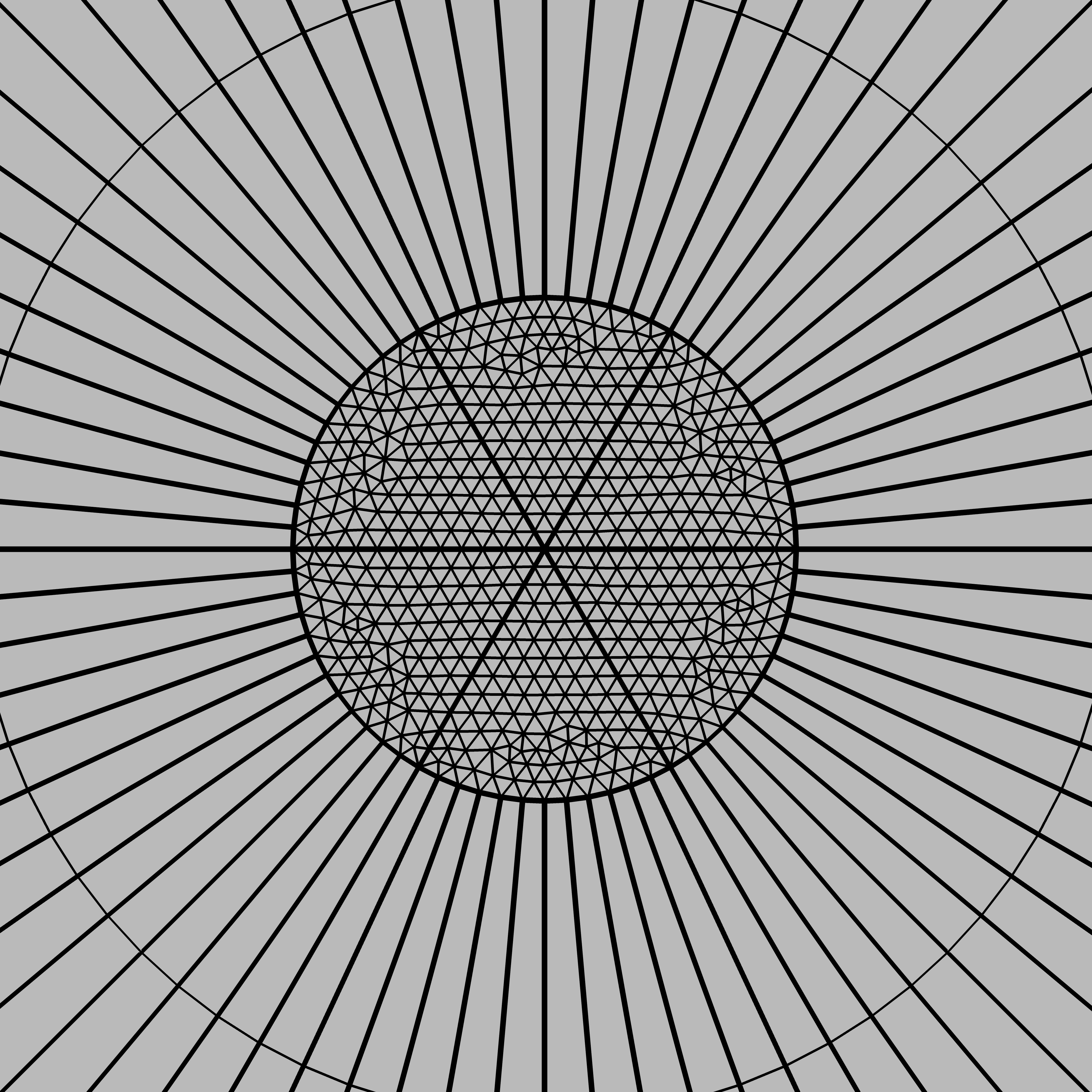}}
        \vfill
    \end{minipage}}
\caption{Domain of the fluid sample model: (a) top view, (b) side view and (c) detail of the top/bottom wall inner region.}
\label{fig:Mesh}
\end{figure}

The mesh was constructed in COMSOL by defining it first on the top and outer walls and then using a swept mesh operation to mesh the whole volume. The whole revolution was divided into $(\Delta \theta=)5^\circ$ slices, and the outer wall (fluid/air interface) was defined by partitioning the gap height, along the $z$ direction, into 20 equal-sized elements. The top wall was divided into two sections, a structured outer region and an unstructured inner one, with a threshold diameter of $R_\mathrm{thr.}=0.5$ mm. On the outer region ($R_\mathrm{thr.}\leq r \leq R$) the radial direction was divided into 30 elements, being progressively refined as we approach the outer rim of the geometry ($r\rightarrow R$) to improve the resolution where the maximum gradients are expected (arithmetic progression with growth factor of 5). The inner region ($0 \leq r \leq R_\mathrm{thr.}$) was constructed by dividing it into angular sections of 60$^\circ$ and filled with small triangular elements (maximum and minimum sizes of $R_\mathrm{thr.}\Delta \theta$ and $R_\mathrm{thr.}\Delta \theta/10$, respectively, with maximum element growth rate of 1.2). This inner/outer division was necessary to prevent elements with collapsed faces at $r=0$. The employed mesh is showcased in Figure \ref{fig:Mesh} and it was composed of 65160 elements, 43200 hexahedral elements in the outer region and 21960 prisms near the centre. More information regarding the mesh construction and analysis is given as supplementary material.

\begin{figure}[htp]
\centering
\includegraphics[width=.7\linewidth]{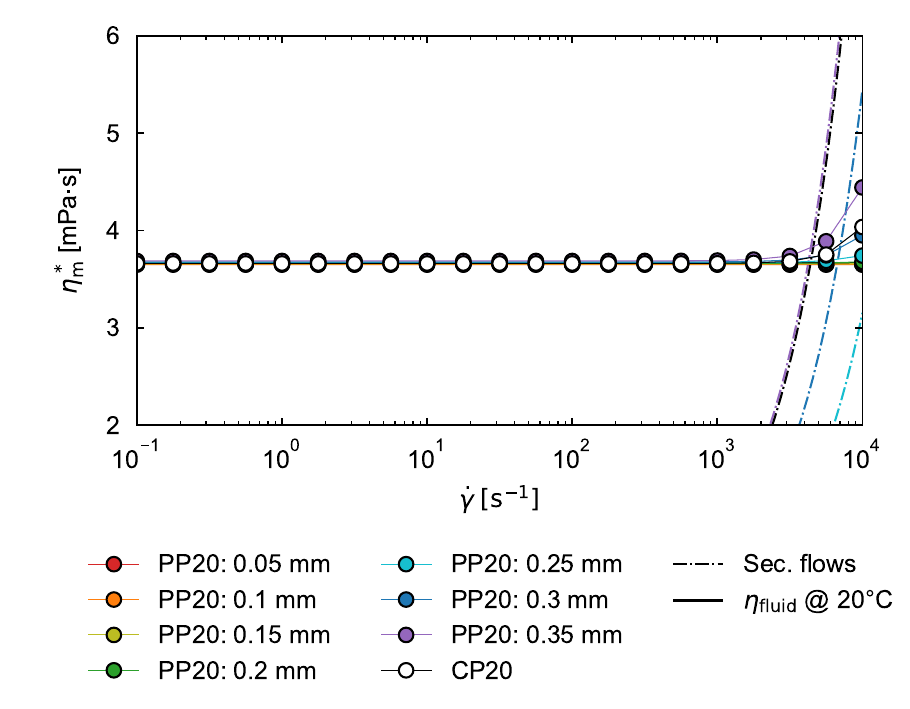}
\caption{Viscosity curves of the Cal. Oil obtained numerically with perfectly parallel ($\varphi = 0$) PP20 ($0.05 \leq h_\mathrm{c} \leq 0.35$ mm) and CP20. The chosen mesh configuration is able to return a constant numerical viscosity equal to the model input value (error below 1\%), for both geometries and all tested PP20 gap. The onset of secondary flows at high shear is also well described and is effectively predicted by the corresponding limit which is also plotted.}
\label{fig:Flow curves - numerical parallel}
\end{figure}

The models were validated by verifying the velocity profiles obtained with perfectly parallel geometries ($\varphi = 0$) at a shear rate range $10^{-1}\leq\dot\gamma \leq 10^{4}$ s\textsuperscript{-1}. The profiles never deviated from the expected Couette shape except at high shear rates when secondary flows onset, which was also accurately captured by the models (data not shown here). The viscosity "measured" by our modelled rheometer was calculated, according to Equation \ref{eq:viscosity}, by evaluating the torque, which was computed by integrating the wall shear stress on the top, rotating wall:
\begin{equation}
    \mathrm{T} = \iint_S\,\tau_{\mathrm{z}\theta}\,r\,dS\,.
    \label{eq:numeric_torque}
\end{equation}
Figure \ref{fig:Flow curves - numerical parallel} shows the viscosity curves "measured" numerically ($\eta_\mathrm{m}^*$) for several PP20 gaps ($0.05\leq h_\mathrm{c}\leq 0.35$ mm) and the CP20. The models returned a constant measured viscosity equal to the defined fluid viscosity (error below 1\%) except when secondary flows onset at high shear, which led to a viscosity overestimation well predicted by the secondary flow limit.

\begin{figure}[htp]
\centering
\includegraphics[width=\linewidth]{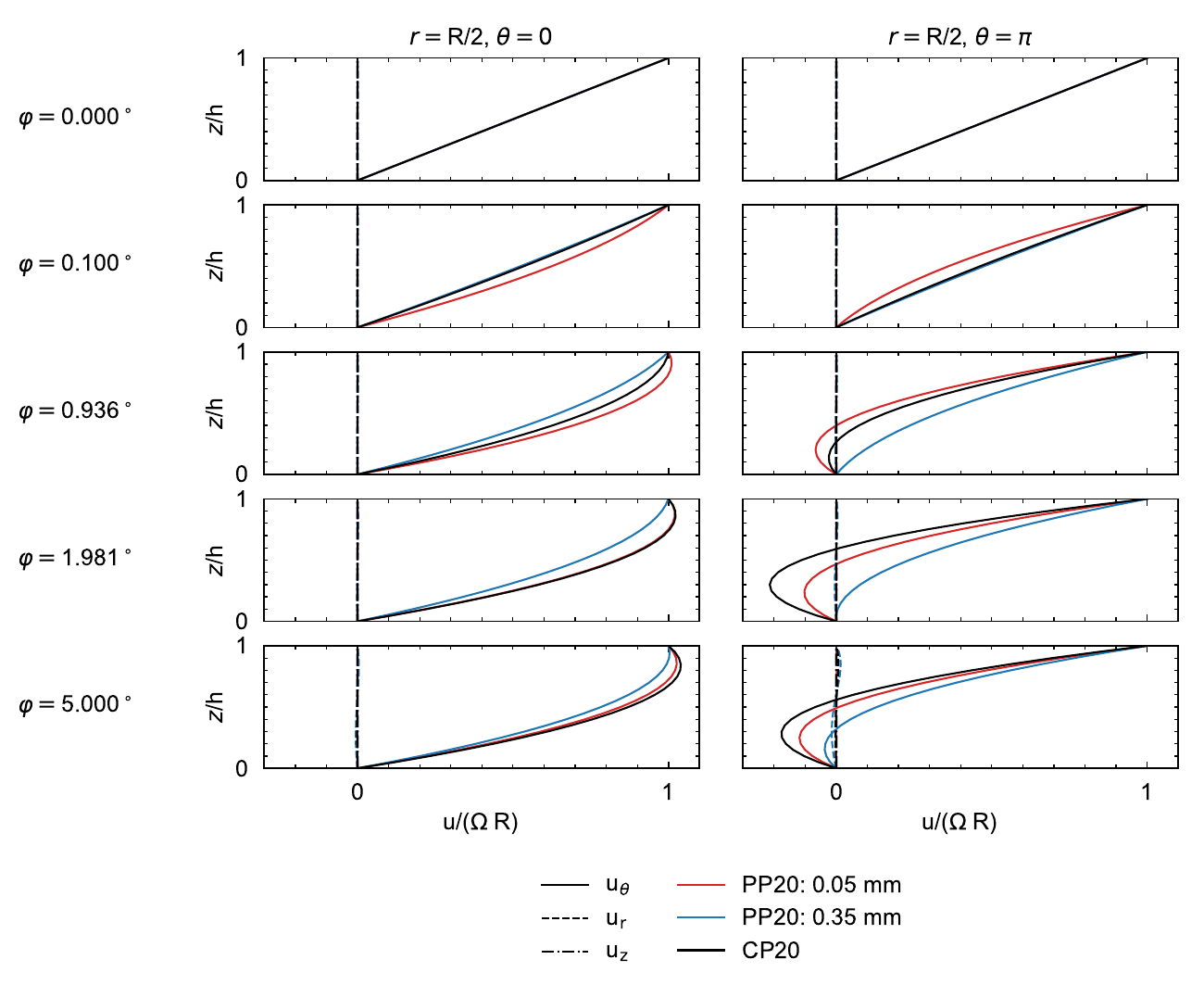}
\caption{Velocity profiles obtained from the modelled flow of the Cal. Oil, along opposing vertical lines: $\theta=0$ and $\pi$ for $r=R/2$, at several inclination angles ($\varphi = 0, 0.1, 0.936, 1.981, 5^\circ$) in the PP20 ($h_\mathrm{c}=0.05$ and $0.35$ mm) and the CP20. The aggravation of the geometry non-parallelism results in significant divergences from the canonical Couette profile.}
\label{fig:Profiles - tilted}
\end{figure}

Having evaluated the model's capability, we introduce the non-parallelism of the geometries by varying the inclination angle of the bottom plate, $\varphi$. To compare with our experimental data, the shear rate was fixed at $\dot\gamma = 115$ s\textsuperscript{-1} (data point within the experimental window) and twenty values for the inclination angle were tested, ten lesser than the cone angle, $0.1^\circ\leq \varphi \leq \beta$, and ten larger, $\beta \leq \varphi \leq 5^\circ$. In Figure \ref{fig:Profiles - tilted} are shown the velocity profiles (velocity components: $u_\mathrm{r}$, $u_\theta$ and $u_\mathrm{z}$) at $r=R/2$ in opposite sides of the inclination ($\theta=0$ and $\theta = \pi$) for five inclination angles: $\varphi = 0, 0.1, 0.936, 1.981$ and $5^\circ$, on both geometries (only two PP20 gap heights are shown: $h_\mathrm{c}=0.05$ and $0.35$ mm).

\begin{figure}[htp]
\centering
\includegraphics[width=\linewidth]{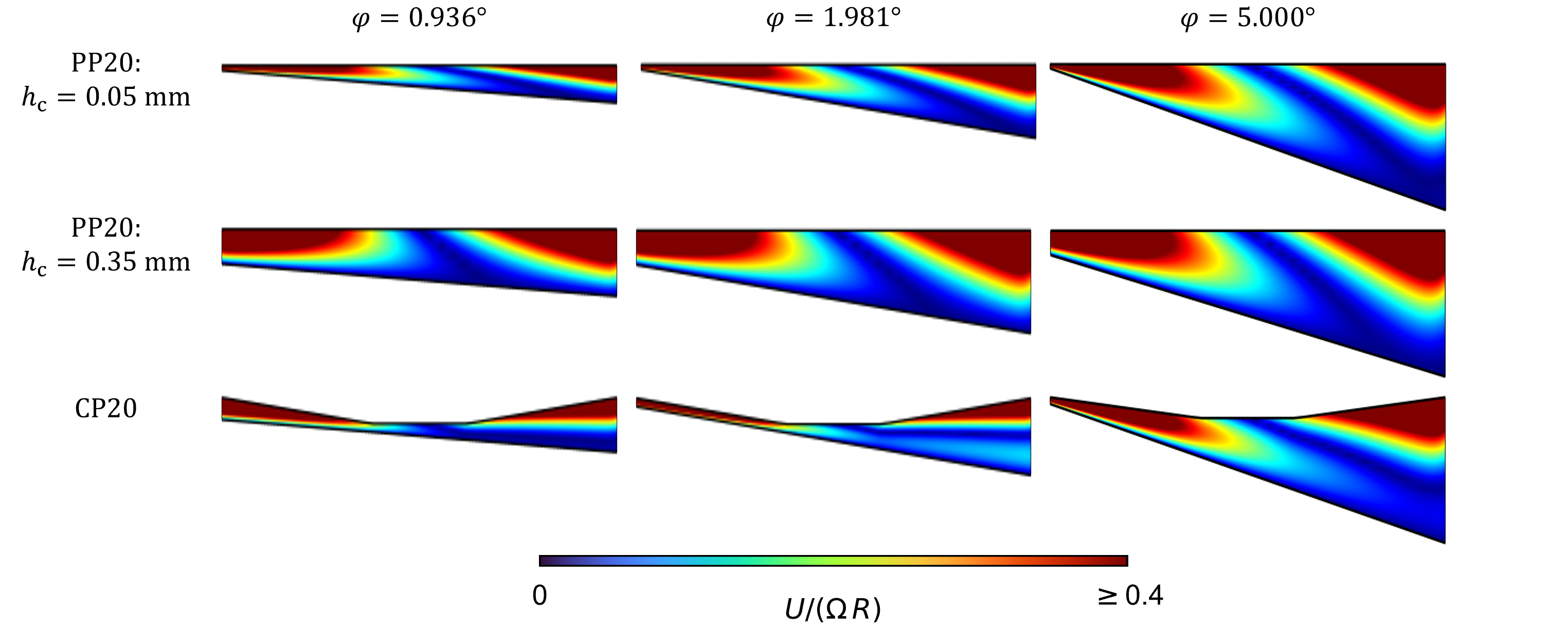}
\caption{Velocity magnitude fields ($U = \sqrt{u_\theta^2+u_r^2+u_z^2}$) gathered at the vertical plane $\theta=0$ on either geometry model and two PP gap heights ($h_\mathrm{c} = 0.05$ and $0.35$ mm) for bottom plate inclinations of $\varphi=0.936^\circ, 1.981^\circ$ and $5.000^\circ$. The colorbar range was limited to 40\% of the data range ($0\leq U\leq 0.4\;\Omega R$) and the plots were stretched vertically (stretching factor of 5) to facilitate visualisation. The bottom plate inclination induces a tilt of the rotational axis of the flow and evidence of a counter-rotating region is observed in the CP20 for large inclinations.}
\label{fig:Velocity_fields}
\end{figure}

\begin{figure}[htp]
\centering
\includegraphics[width=\linewidth]{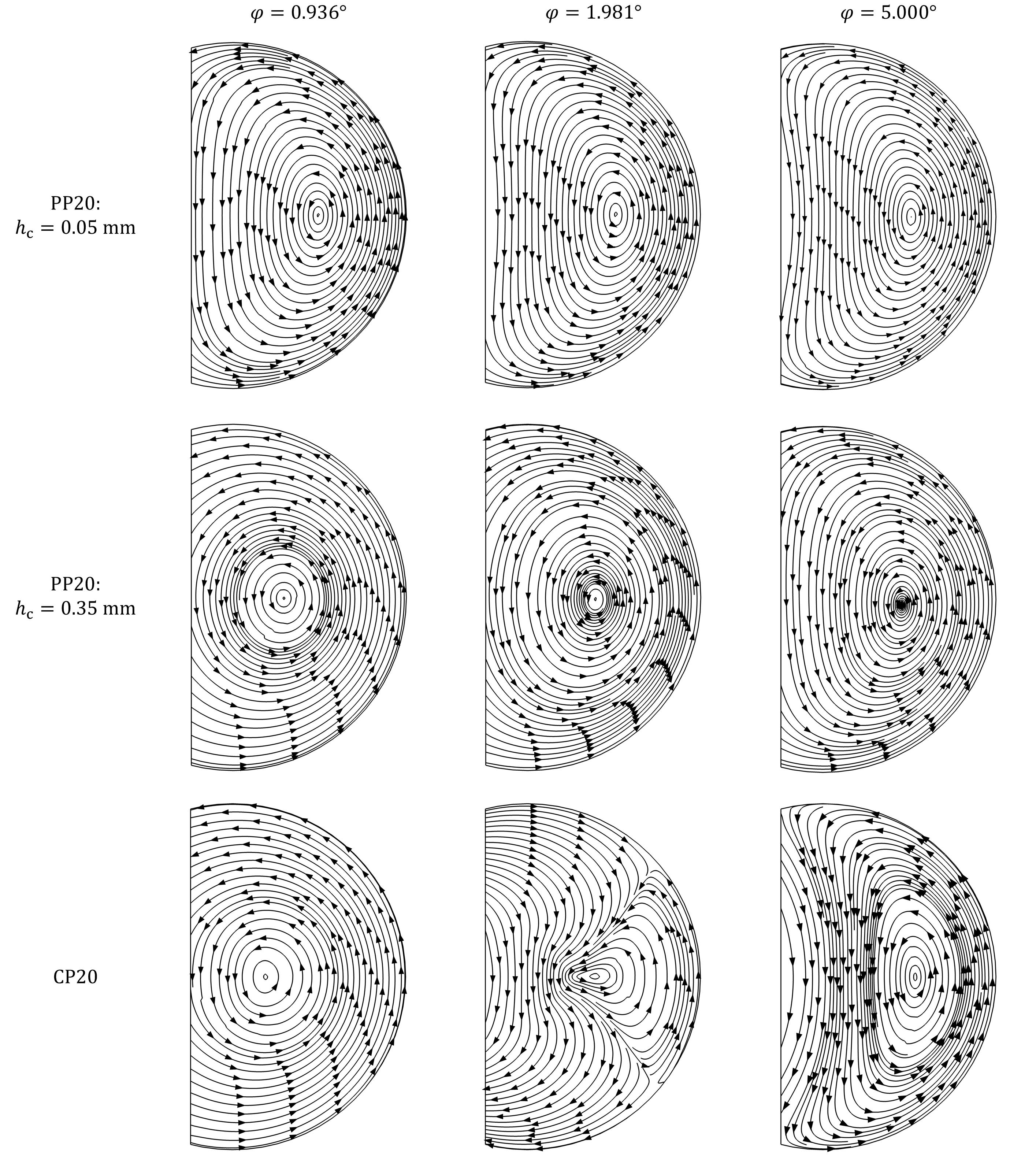}
\caption{Streamlines gathered at an horizontal plane (crossing the inclined bottom plate at $r=R_\mathrm{t}$ and $\theta = 0$) on either geometry model and two PP gap heights ($h_\mathrm{c} = 0.05$ and $0.35$ mm) for bottom plate inclinations of $\varphi=0.936^\circ, 1.981^\circ$ and $5.000^\circ$. Imposed rotational velocity on the top wall is counter-clockwise. A recirculating region is observed in the CP geometry for large bottom plate inclinations.}
\label{fig:Streamlines}
\end{figure}

The inclination of the bottom plate leads to divergences from the canonical Couette flow (Figure \ref{fig:Profiles - tilted}). Whereas the CP20 and the larger PP20 gap appear relatively unaffected by the lesser tested inclination ($\varphi = 0.1^\circ$), the smaller PP20 gap is more sensitive due to the inclination's increased relative impact. Despite the non-parallelism of PP geometries provoking only an increase of the local gap, the angular velocity profiles on either side of the geometry are curved differently. Notably, aggravating the non-parallelism leads to interesting velocity profiles. On the one hand, at the lesser local gaps ($\theta=0$) there seems to be a velocity overshoot, noticed first for the lesser PP20 gap and then for the CP20. On the opposite location ($\theta=\pi$), negative rotational velocities are seen near the bottom wall, becoming very prominent particularly for the CP20 when the inclination matches the cone angle ($\varphi=\beta=1.981^\circ$). To further analyse the flow, Figures \ref{fig:Velocity_fields} and \ref{fig:Streamlines} show, respectively, velocity magnitude fields (at the vertical plane $\theta = 0$) and streamlines (at the horizontal plane crossing the inclined bottom plate at $r=R_\mathrm{t}$, $\theta = 0$) for either geometry model (and both PP20 gaps: $h_\mathrm{c} = 0.05$ and $0.35$ mm), and three bottom plate inclinations.

It seems there are fundamental differences between the flow on either non-parallel geometry. Regarding the PP20, from the velocity magnitude fields we can see the inclination of the bottom plate leads to a tilt of the flow's axis of rotation, being more pronounced for smaller gaps. This explains the negative angular velocities displayed on the profiles (Figure \ref{fig:Profiles - tilted}) as they were gathered taking into account the geometry's axis and not the flow's. On the other hand, the flow behaviour on the non-parallel CP is more complex and, for inclinations around and larger than $\beta$, near the bottom of the geometry the flow counters the rotation imposed on the top wall (clearly displayed by the streamlines). This phenomenon should be due to the contraction/expansion provoked by the inclination, which has a maximum strangulation when it matches the cone angle. Increasing the inclination past this point shifts the most narrow section from the truncation radius to the outer radius, widening the contraction and, therefore, beginning to slowly dissipate the counter-rotating region.

\begin{figure}[htp]
\centering
\includegraphics[width=.7\linewidth]{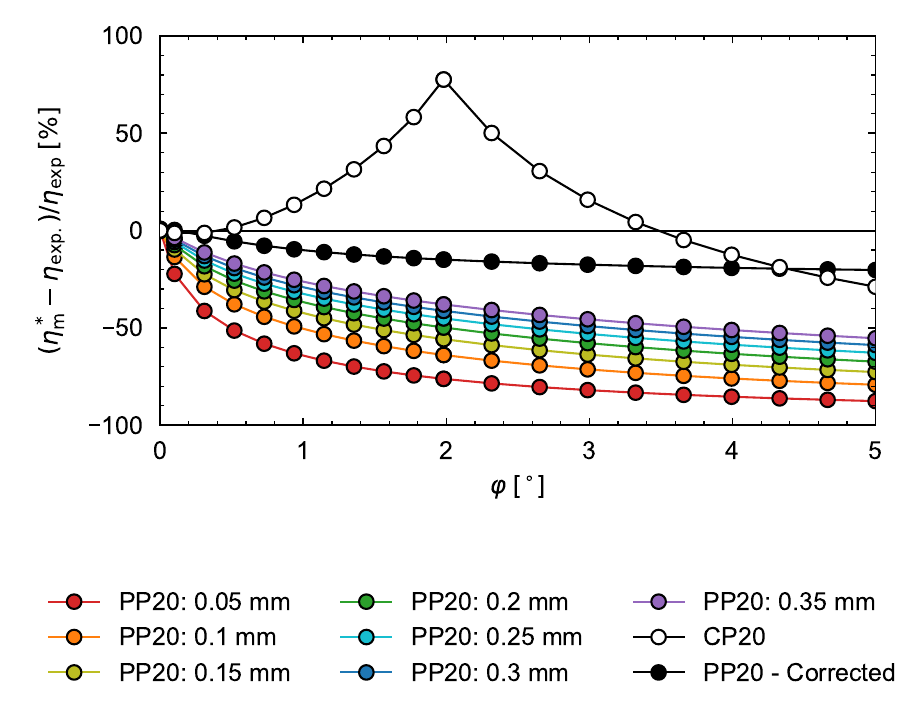}
\caption{Modelled Cal. Oil numerical viscosity ($\eta_\mathrm{m}^*$, relative to the input value, $\eta_\mathrm{exp}$, given in Table \ref{tab:fluids}) with varying inclination of the bottom plate ($0\leq\varphi\leq5^\circ$) on the PP20 ($0.05 \leq h_\mathrm{c} \leq 0.35$ mm) and CP20 geometries. The geometry non-parallelism results in a measured-viscosity decrease with diminishing PP gap, which can be corrected relatively well through the gap-error formulation for small inclinations. With the CP, however, a measured-viscosity overshoot is noticed, reaching a maximum for a critical inclination equal to the cone angle.}
\label{fig:Viscosity - inclination}
\end{figure}

Figure \ref{fig:Viscosity - inclination} shows the viscosity "measured" by our modelled rheometer ($\eta_\mathrm{m}^*$, relative to the input Cal. Oil viscosity, $\eta_\mathrm{exp}$, given in Table \ref{tab:fluids}) with the PP20 ($0.05 \leq h_\mathrm{c} \leq 0.35$ mm) and CP20 geometries with varying inclination ($0\leq\varphi\leq5^\circ$). With the PP20, the measured viscosity decreases as the inclination is increased, tending to zero as $\varphi \to 90^\circ$ with a decreasing rate. Despite seemingly behaving as a usual gap-error on perfectly-parallel geometries, the gap-error formulation returned viscosity estimates lower than the models' input value, possibly due to the flow alterations displayed in Figures \ref{fig:Velocity_fields} and \ref{fig:Streamlines}. This can also explain the observed underestimation of the corrected experimental data (Figure \ref{fig:Gap-error_correction}). Nonetheless, relatively acceptable errors (<10\%) are still achievable through the gap-error formulation for small inclinations $\varphi\lesssim1^\circ$. 

With the CP20, the measured viscosity is initially increased until reaching a maximum when the inclination matches the cone angle, about 77\% higher than the actual viscosity. This should be due to the progressive "pinching" of the contraction/expansion. For $\varphi\leq\beta$, on one side of the geometry ($\theta = 0$) the increase in inclination reduces the flow depth, leading to enhanced velocity gradients and, therefore, larger torque contributions. Whereas on the other side ($\theta = \pi$) there is an increase in flow depth, but the counter-rotating region is formed near the bottom wall, allowing for the maintenance of significant velocity gradients and, thus, the overall torque is increased. Increasing the inclination past the cone angle ($\varphi>\beta$), the contraction is progressively reduced, which naturally leads to, on one hand, a reduction of the velocity gradients at the constricted section and, on the other hand, to the dissipation of the recirculation region, which leads to a decrease of the measured viscosity.

These results corroborate the experimental observation of the PP20-measured viscosity decreasing with gap reduction and the slight increase in CP20-measured viscosity. A standard gap-error on perfectly parallel geometries (arising, for example, from miss-judged zero-gap, due to air squeeze or other phenomena) would only lead to a viscosity decrease (this was also evaluated for our models and is presented as supplementary material). Thus, according to the previous experimental data we can infer that the bottom MP is likely slightly inclined. Fitting the numerical viscosity, a gap-error of $\varepsilon=22.8$ \textmu m should be estimated from an inclination between $0.10^\circ<\varphi<0.31^\circ$. However, according to Figure \ref{fig:Viscosity - inclination}, the CP-viscosity increase should not be noticeable for such a slight inclination, which disagrees with our experimental observations. Other surface defects could have a similar effect on experimental results; however, the issue may lie not with the quality of the plate's surface but with its coupling to the magnetorheological cell. Looking again at Figure \ref{fig:Setup}(b2), we can observe that the MP is fixed to the cell via two screws and either dimensional issues with the screws themselves or surface defects on the bottom of the plate could be responsible for this inclination.

Slight geometry non-parallelism can be predicted and corrected fairly well for PP geometries through the gap-error formulation and viscosity data at multiple gap heights. However, as far as we know, there is no equivalent approach for CP geometries. The flow dynamics can become relatively complex, and the measurements can return unexpected results as we usually assume geometry non-parallelism to result similarly to a general gap error. From this numerical work, we can gather that reasonably acceptable errors (<10\%) can be obtained for standard measurements with a CP20 geometry for small inclinations, $\varphi\lesssim1^\circ$. For $3^\circ\lesssim\varphi\lesssim4^\circ$ the CP also returns small errors, but the divergence from the canonical flow can significantly impact the behaviour of the used sample. Until now, we have focused on single-phase Newtonian fluids, but measurements of more complex samples may be affected differently. In the following Sections, we shall discuss the impacts of geometry non-parallelism on magnetorheological measurements.

\subsection{Experimental magnetorheological measurements}
\begin{figure}[htp]
\centering
\includegraphics[width=\linewidth]{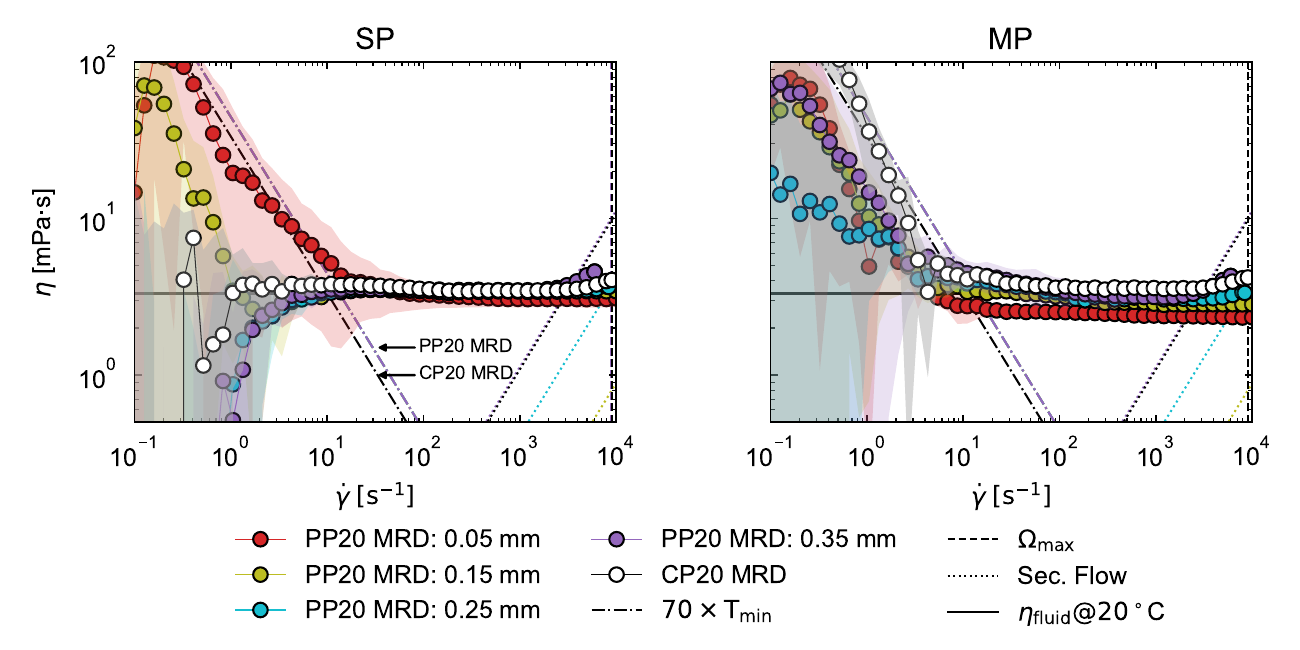}
\caption{Viscosity curves obtained with the unseeded Newtonian blood analogue (NBa), on either bottom plate: (left) SP and (right) MP. Data gathered with the CP20 MRD and PP20 MRD (gap heights between 0.05 and 0.35 mm). No magnetic field was applied ($B=0$). On the MP, the same PP-measured viscosity gap-dependence is noticed, as is the slight CP overshoot.}
\label{fig:NBa_Flow_curves}
\end{figure}

Measurements were performed to evaluate the setup capability of generating a measurable magnetorheological response and how the used geometry impacts the results. To this end we employed the Newtonian blood analogue, NBa (composition and expected density and viscosity in Table \ref{tab:fluids}). First, the analogue's viscosity curves were obtained without particles and no magnetic field application, using both geometries and bottom plates. Tests were conducted with four different gap heights for the PP, ranging from 0.05 to 0.35 mm. The resulting viscosity curves are illustrated in Figure \ref{fig:NBa_Flow_curves}. Again, the data is similarly affected by the gap-error, showing decreased PP-measured viscosity with gap height reduction and increased CP-measured viscosity. On the other hand, the NBa seems more sensitive to low-shear issues than the Cal. Oil, particularly for larger gaps on the MP, which could be due to the analogue's more significant surface tension.

\begin{figure}[htp]
\centering
\includegraphics[width=\linewidth]{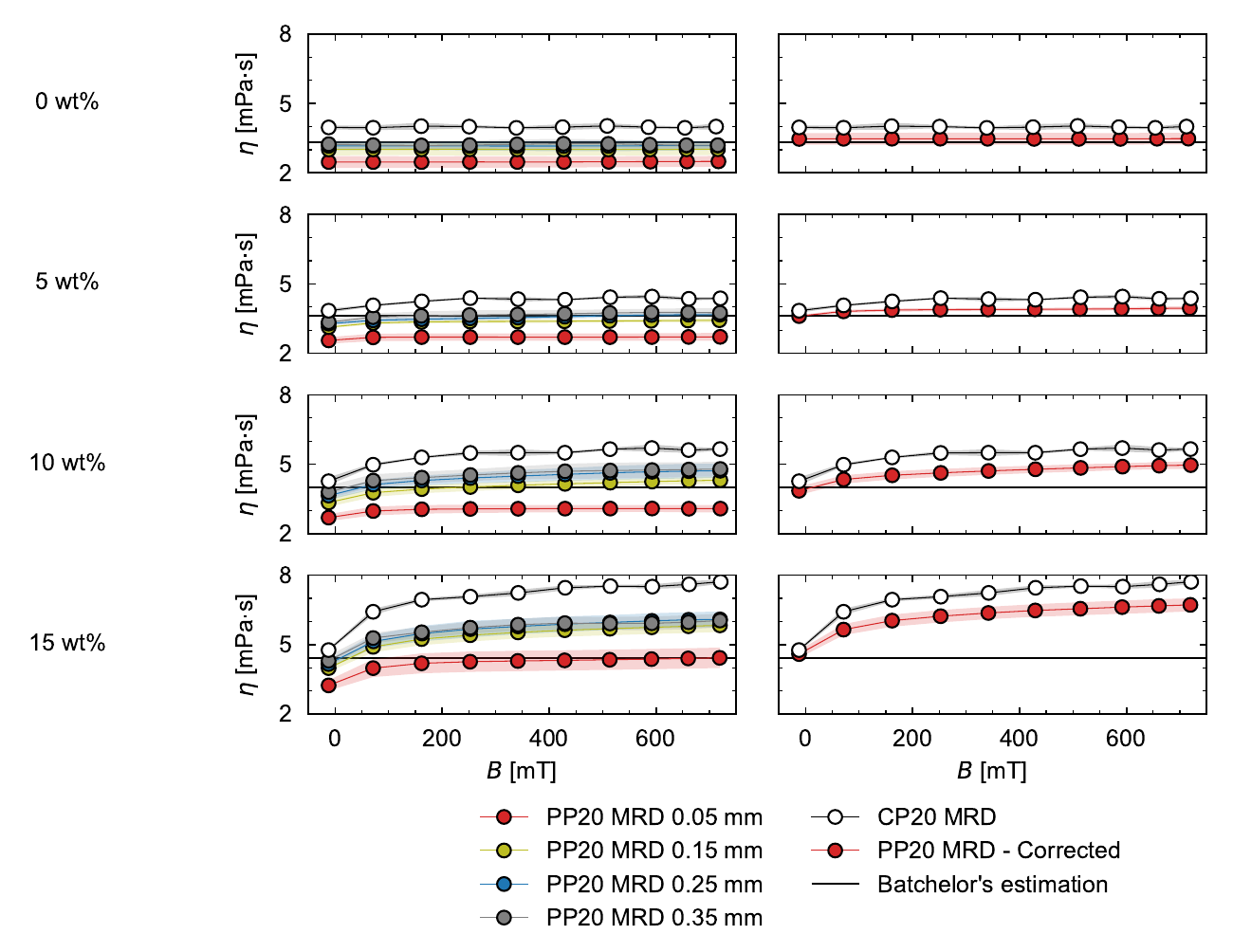}
\caption{Viscosity data (at $\dot\gamma = 500$ s\textsuperscript{-1}) of the NBa seeded with magnetic particles at different mass concentrations (0, 5, 10 and 15 wt\%), on the MP with varying magnetic field density ($B\leq720$ mT). Data gathered with the CP20 MRD and PP20 MRD (gap heights between 0.05 and 0.35 mm). In the right column, the PP20 data gathered at $h_\mathrm{c}=0.15$ mm is shown corrected for the gap-error ($\varepsilon=22.8$ \textmu m). Horizontal black lines correspond to Batchelor's estimated viscosity of the seeded NBa at 20$^\circ$C (Equation \ref{eq:batchelor}). The application of the external magnetic field results in a clear viscosity increase of the seeded NBa samples, particularly for larger particle concentrations.}
\label{fig:NBa - Magnetic}
\end{figure}

The magnetorheological steady shear measurements were conducted with the NBa seeded with three concentrations of the M270 particles: 5, 10 and 15 wt\% (volume fractions of $\phi \approx 3.4, 6.9$ and 10.6 vol\%). The shear rate was kept constant at $\dot\gamma=500$ s\textsuperscript{-1} (within the experimental window of the unseeded NBa) and the magnetic field intensity was varied up to $B\leq720$ mT, following the procedure described in Subsection \ref{subsec:Methods_magnetorheological}. The measurements were conducted on the MP, with both geometries (PP20 MRD and CP20 MRD). With the PP20 MRD, the fluid viscosity was measured at different gap heights: $0.05 \leq h_\mathrm{c} \leq 0.35$ mm, (left column of Figure \ref{fig:NBa - Magnetic}), and the data gathered at an intermediate gap, $h_\mathrm{c}=0.15$ mm, was corrected for the gap-error, $\varepsilon=22.8$ \textmu m (right column of Figure \ref{fig:NBa - Magnetic}).

Before discussing the results, we address some possible experimental issues. Evaluating the Peclet, Reynolds and Stokes numbers, the results to Equations \ref{eq:Peclet}-\ref{eq:Stokes} are, respectively: $1/Pe\approx4.7\times10^{-5}$, $Re_\mathrm{p}\approx3.1\times10^{-4}$ and $St\approx1.0\times10^{-4}$, which are $\ll1$, meaning we can disregard Brownian and inertial effects. Regarding particle sedimentation, the settling velocity in a dilute system can be given by:
\begin{equation}
    V_\mathrm{s} = (1-6.55\,\phi)\,\frac{(d_\mathrm{p}/2)^2\,\Delta\rho\, g}{18\eta}\,,
\end{equation}
where the term $(1-6.55\,\phi)$ accounts for the hindered backflow of the continuous phase due to the presence of the particles, $\Delta\rho$ is the density difference between dispersed and continuous phases and $g$ is the gravitational acceleration\citep{batchelor1972}. The resulting sedimentation velocities are, for 5, 10 and 15 wt\% concentrations: $V_\mathrm{s} \approx 0.55, 0.39$ and $0.22$ \textmu m/s, respectively, which may be significant given our small gap heights, even if the total measurement time was relatively short (200 s). However, because the magnetic field is aligned with the gap, the magnetic dipole forces will counter the gravitational effects, further hindering the particle sedimentation. 

Looking at the results obtained without magnetic field application, $B = 0$ there is an increase in viscosity with particle concentration, which is reasonably predicted by Batchelor's expression\citep{batchelor1977} (shown in Figure \ref{fig:NBa - Magnetic} as horizontal black lines):
\begin{equation}
    \eta^* = \eta\;(1+2.5\phi+6.2\phi^2)\,.
    \label{eq:batchelor}
\end{equation}

Regarding the magnetic field effects, the results present the expected response\citep{felt1996}: an increase in viscosity is noted with enhanced magnetic field density, which is more significant the larger the particle concentration. This is because the particle-chains, formed from the magnetic-induced dipole interactions, are aligned (with the field) perpendicular to the flow direction, hindering it, provoking an additional torque and, therefore, an increase in measured viscosity. The response becomes progressively less sensitive to the magnetic field alterations as its intensity is increased, as expected from the particles' magnetisation curve\citep{grob2018}, eventually reaching a saturated state for $B\gtrsim500$ mT. There is a discrepancy between the PP20 and CP20 results, with the latter always presenting larger measured viscosities, and may seem dependent on the particle concentration and the magnetic field density. Nevertheless, this could be the same effect of the bottom plate inclination as previously discussed, because the relative error between the two geometries remained relatively constant ($(\eta_\mathrm{CP}-\eta_\mathrm{PP})/\eta_\mathrm{CP} \approx 13.5 \%$), it just becomes visually prominent due to the viscosity increase with magnetorheological enhancement. 

It is worth mentioning that because we employed smooth geometries, our data is likely affected by apparent slip of the dispersed phase which would significantly reduced the measured magnetorheological response\citep{vicente2004,buscall2010}. As such, these results serve only to verify that the employed setup is able to generate a significant magnetorheological response and to compare between the discussed geometries. We note that inhibiting slip, either through roughened or detailed geometries, should return a much stronger magnetorheological behaviour with measured viscosities perhaps up to orders of magnitude larger. 

Lastly, and on a side note, heating effects from the magnetic field generation were observed from a very slight temperature rise at the largest field densities, but it should not have significantly affected the results ($<0.3^\circ$C).

\subsection{Numerical analysis of the non-parallelism on magnetorheological measurements}
As the SP is not applicable for magnetorheology, decoupling the effects of the MP's inclination on this type of measurement was not possible. Thus, the flow of magnetised particles seeded in the NBa was evaluated numerically using the rheometer models. The experimental procedure would ideally be replicated to assess the impact of non-parallelism on bulk viscosity measurements and particle chain dynamics. However, the required number of modelled particles results in extensive simulations that are, unfortunately, beyond our current computational capabilities. 

Taking a different approach, the effects of geometry non-parallelism on particle-chain behaviour can be assessed by evaluating only a few chains with a small number of particles placed strategically in the flow. We placed particle chains at half the geometry radius: $r = R/2$ at four equally-distanced angular positions: $\theta \in [0,\,\pi/2,\,\pi,\,3\pi/2]$. To have coherency between the geometries, the PP20 commanded gap was defined as: $h_\mathrm{c}^\mathrm{PP20} = h^\mathrm{CP20}|_{R/2} = (R/2)\tan(\beta)$.

COMSOL's \textit{particle tracing for fluid flow} module was used. The forces considered to be acting on the particles were the viscous drag, due to the fluid flow, and dipole-dipole interactions, due to particle magnetisation. Particle sedimentation has already been discussed in the previous Section and again addressing Brownian motion, \citet{melle2003} defined an adimensional ratio between magnetic and thermal energies:
\begin{equation}
    \lambda = \frac{\mu_0\,\mu_\mathrm{f}\,m^2}{16\,\pi\,(d_\mathrm{p}/2)^3\,k_B\,T}\,,
\end{equation}
where $k_B$ is the Boltzmann constant and $T$ is the temperature, which we defined as 20$^\circ$C. Using the M270 saturation magnetisation\citep{grob2018}($m_\mathrm{sat.}^\mathrm{M270} = 6.4\,\times\,V_\mathrm{p}\,\rho_\mathrm{p}$ [Am\textsuperscript{2}], where $V_\mathrm{p}$ is the particle volume), we obtain $\lambda \approx 32000\gg1$, meaning the magnetic effects dominate over Brownian motion. 

Viscous drag was computed using COMSOL's in-built Stokes drag force:
\begin{equation}
    \boldsymbol{F_\mathrm{d}} = 3\,\pi\,\eta\,d_\mathrm{p}(\boldsymbol{u_\mathrm{f}}-\boldsymbol{u_\mathrm{p}})\,,
\end{equation}
where $(\boldsymbol{u_\mathrm{f}}-\boldsymbol{u_\mathrm{p}})$ is the particle velocity ($\boldsymbol{u_\mathrm{p}}$) relative to the fluid ($\boldsymbol{u_\mathrm{f}}$). The dipolar forces were computed through particle-particle interactions following the expression given by \citet{melle2003}:
\begin{equation}
    \boldsymbol{F_\mathrm{a,ij}} = \frac{3\,\mu_0\,\mu_\mathrm{f}\,m^2}{4\,\pi\,r_\mathrm{ij}^4}\left[\left(1-5\left(\hat{m}\cdot\hat{r}_\mathrm{ij} \right)^2\right)\hat{r}_\mathrm{ij}+2\left(\hat{m}\cdot\hat{r}_\mathrm{ij} \right)\hat{m}   \right]\,,
\end{equation}
where $\mu_0$ and $\mu_\mathrm{f}$ are the vacuum and fluid permeabilities, respectively (the latter being the relative permeability and considered $\mu_\mathrm{f}=1$). This expression assumes all particles are equally affected by the external field, i.e., their magnetisation is identical and aligned with the magnetic field, which was considered uniform and perpendicular to the flow direction. The magnetic force depends on two vector components: the magnetic moment $\boldsymbol{m}$ and the inter-particle distance\footnote{The inter-particle distance is the vector between particle centres.} $\boldsymbol{r}_\mathrm{ij}$ (with magnitude $m$ and $r_\mathrm{ij}$, and adimensional directions $\hat{m}$ and $\hat{r}_\mathrm{ij}$, respectively).

The particles are considered hard spheres and, to approximate this behaviour, a repulsive excluded-volume force was applied\cite{melle2003,gao2012}:
\begin{equation}
    \boldsymbol{F_\mathrm{r,ij}} = 2\,\frac{3\,\mu_0\,\mu_\mathrm{f}\,m^2}{4\,\pi\,d_\mathrm{p}^4}\exp\left[{-30\left(\frac{r_\mathrm{ij}}{d_\mathrm{p}}-1\right)}\right]\hat{r}_\mathrm{ij}\,.
\end{equation}
The excluded volume force perfectly balances the magnetic attraction when two particles are in mechanical contact ($r = d_\mathrm{p}$) and aligned with the field. Whereas, when the inter-particle distance is $r = 1.1\,d_\mathrm{p}$, the excluded-volume force is approximately 14 times smaller than the magnetic attraction force, not provoking unwanted forces in other surrounding particles\cite{gao2012}. Both these forces were computed as particle-particle interactions, and, therefore, we define a general magnetic force acting on the particles as:
\begin{equation}
    \boldsymbol{F_\mathrm{m,ij}} = \boldsymbol{F_\mathrm{a,ij}} +\boldsymbol{F_\mathrm{r,ij}}\,.
\end{equation}

In truth, the presence of the particles induces changes in the fluid flow, but, despite COMSOL allowing for the coupling of different physics, this would further complicate the model. Therefore, we opted only to evaluate the effects of the fluid on the particle chains and disregard the inverse phenomenon. This approach allows the capture of the general chain dynamics and has been used in similar studies \citep{melle2003}, but the actual magnitude of the simplification is difficult to ascertain as the fluid flow at the particle level is complex. However, we expect that disregarding the flow alterations leads to a decrease in chain stability because, in reality, the surrounding particles hinder the local flow, reducing the overall viscous shear acting on the chains. Furthermore, as we have already simplified the problem to analyse only a few chains, the particle effects on the total flow, and therefore on the total torque measurement, would be negligible and not representative of a real experiment. 

The particle boundary condition forced the particles that collide with the geometry walls or the simplified fluid/air interface to adhere to them with no slip, reducing their velocity to zero. As we are only interested in the behaviour of a general particle chain for a qualitative analysis, we could alter the particle diameter to reduce the required number of particles per chain and their magnetisation, which, when decreased, allows lesser magnetic forces to facilitate convergence. 

\begin{figure}[htp]
\centering
\begin{minipage}[t]{\textwidth}
\centering
(a) Unbroken
\end{minipage}
\begin{minipage}[t]{\textwidth}
\centering
\includegraphics[width=.4\linewidth,trim={0 2cm 2cm 3cm},clip]{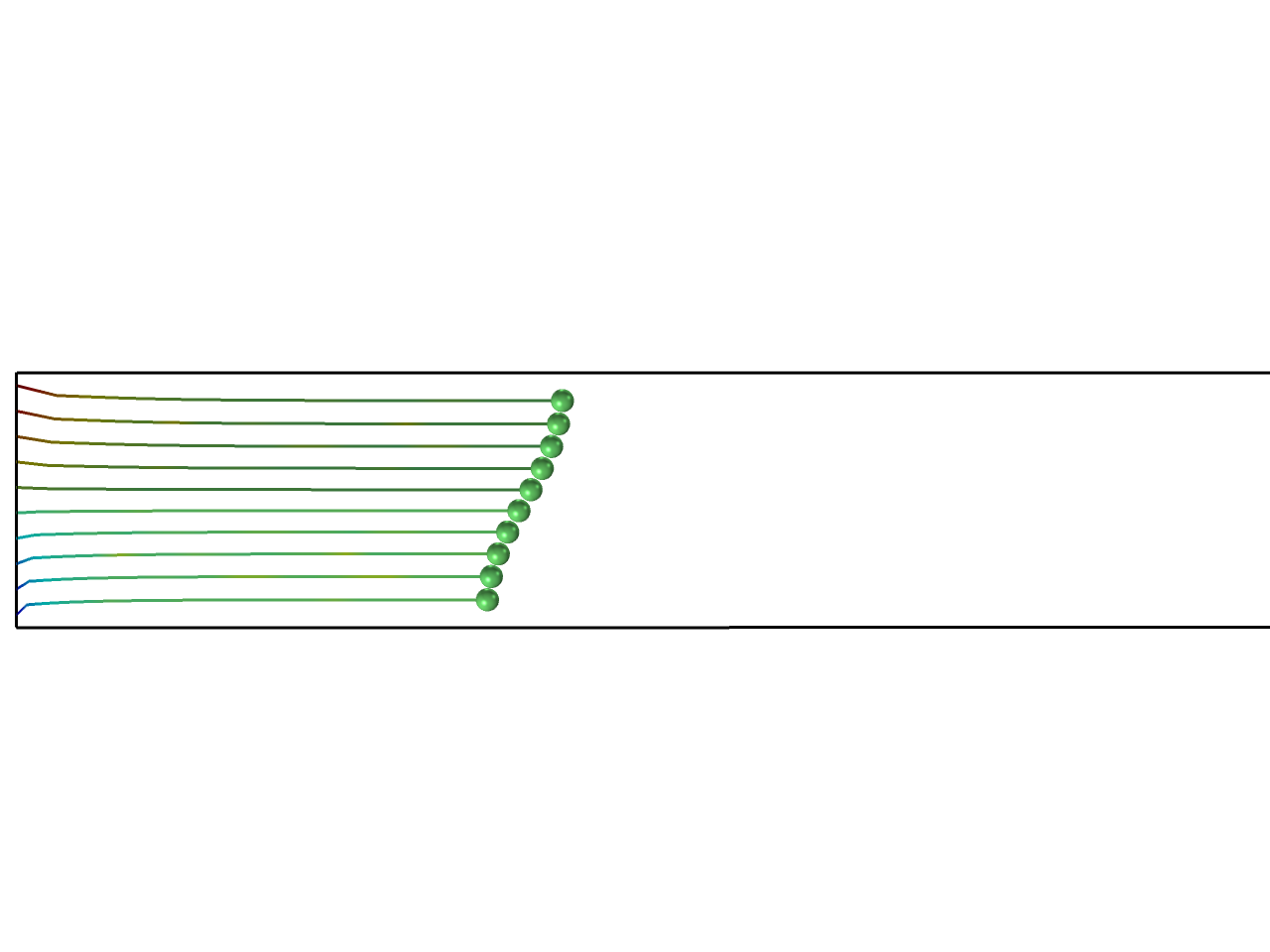}
\end{minipage}
\begin{minipage}[t]{\textwidth}
\centering
(b) Subdivision
\end{minipage}
\begin{minipage}[t]{\textwidth}
\centering
\includegraphics[width=.4\linewidth,trim={0 2cm 2cm 3cm},clip]{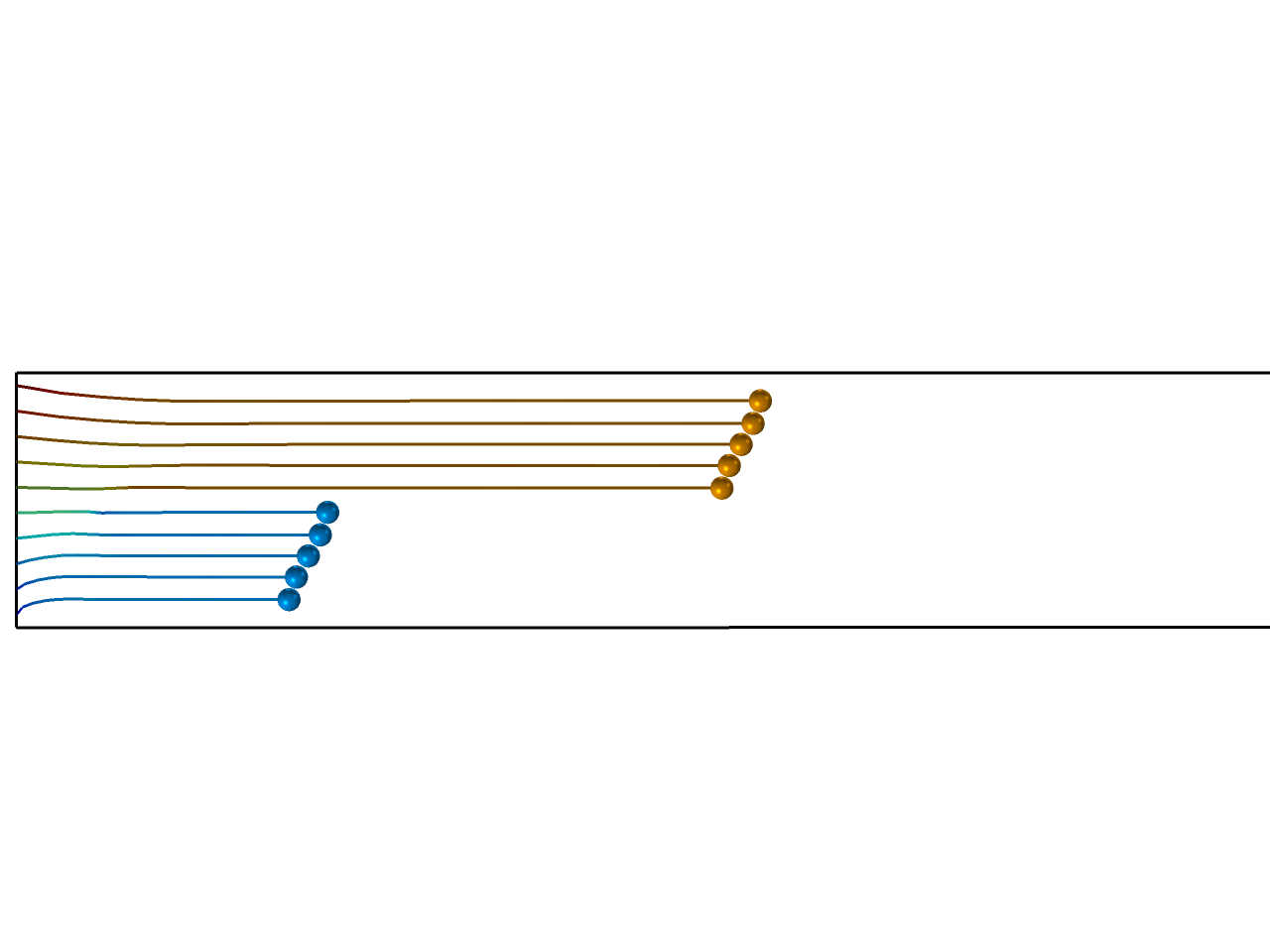}
\includegraphics[width=.4\linewidth,trim={0 2cm 2cm 3cm},clip]{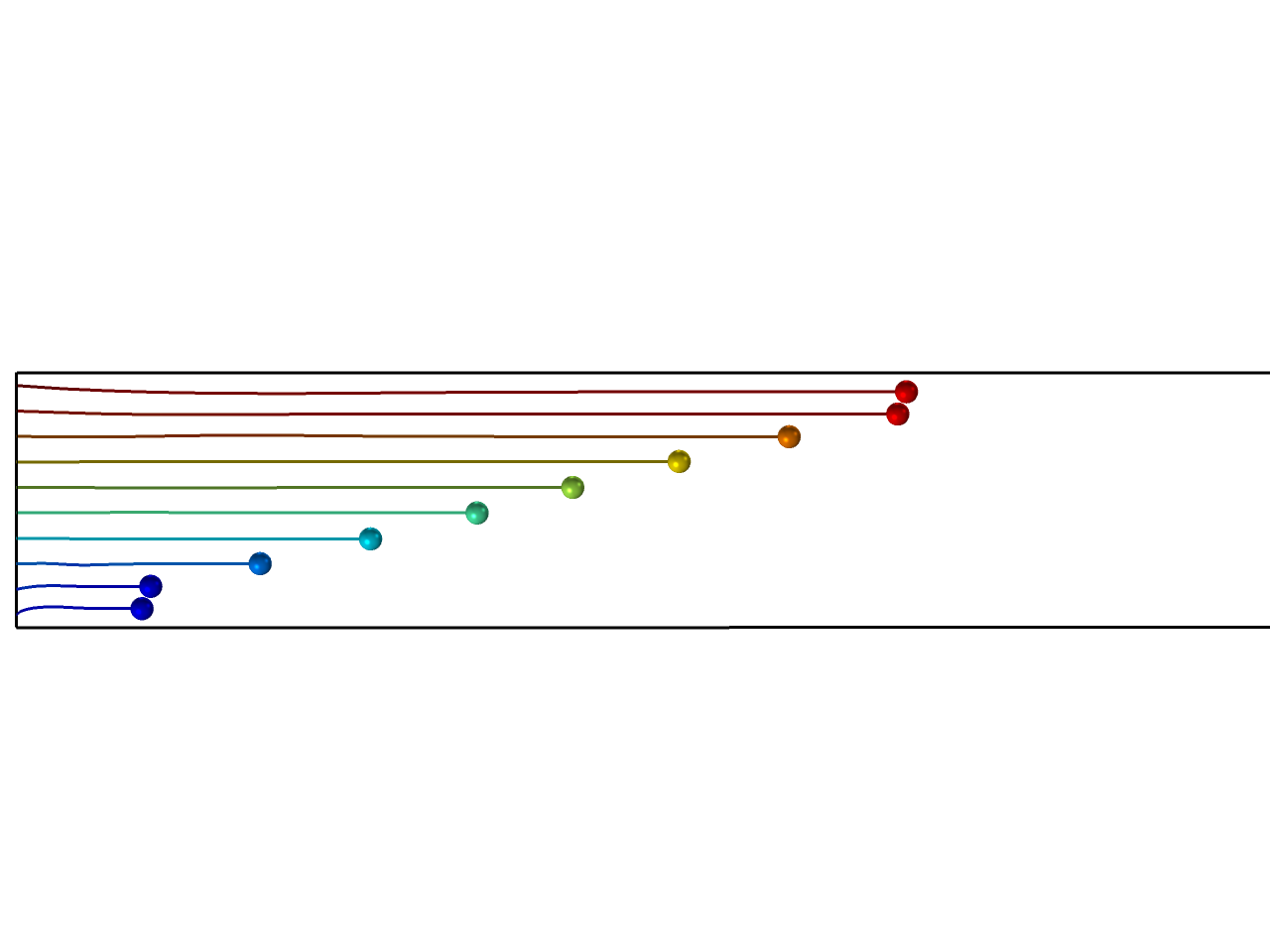}
\end{minipage}
\begin{minipage}[t]{\textwidth}
\centering
(c) Breakdown
\end{minipage}
\begin{minipage}[t]{\textwidth}
\centering
\includegraphics[width=.4\linewidth,trim={0 2cm 2cm 3cm},clip]{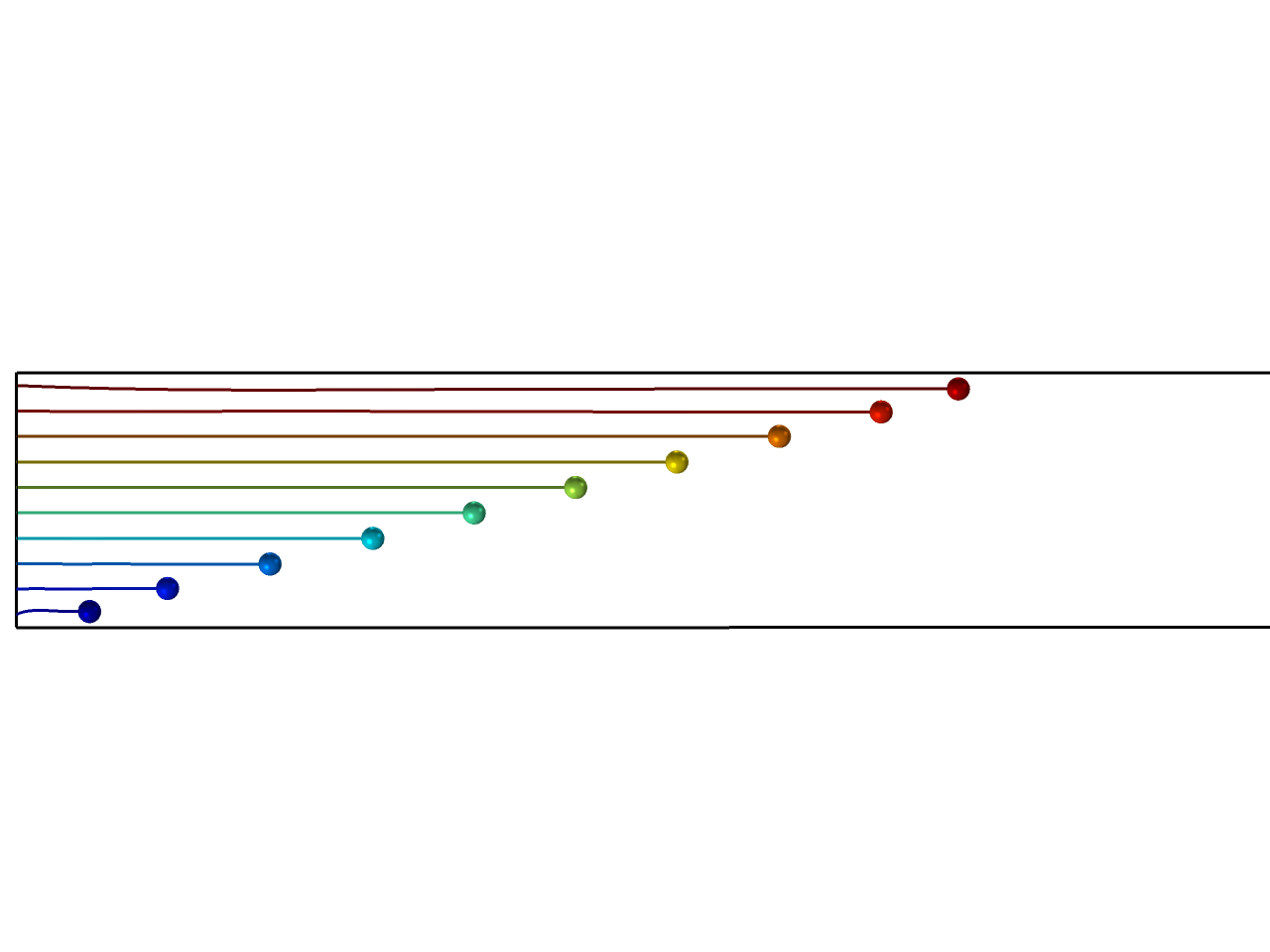}
\end{minipage}
\begin{minipage}[t]{\textwidth}
\centering
\includegraphics[width=\linewidth]{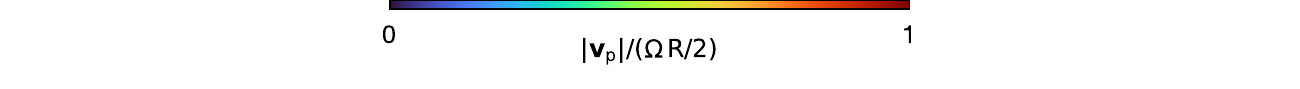}
\end{minipage}
\caption{Magnetised-particle chain ($d_\mathrm{p} = 16$ \textmu m, $n_\mathrm{p} = 10$) on the CP geometry model (without bottom plate inclination) with imposed shear rate $\dot\gamma = 1$ s\textsuperscript{-1} with varying magnetisation factor: (a) $C_\mathrm{m} = 5$, (b) $C_\mathrm{m} =$ (left) 10 and (right) 20, (c) $C_\mathrm{m} = 30$. The particles are initially placed in a vertical uniform distribution (vertical line segment on the left, of length, $R/2\,\tan(\beta)$) and the shown images were taken at $t = 4$ s. Flow is left to right.}
\label{fig:Chain_division_examples}
\end{figure}

\begin{figure}[htp]
\centering
\includegraphics[width=.7\linewidth]{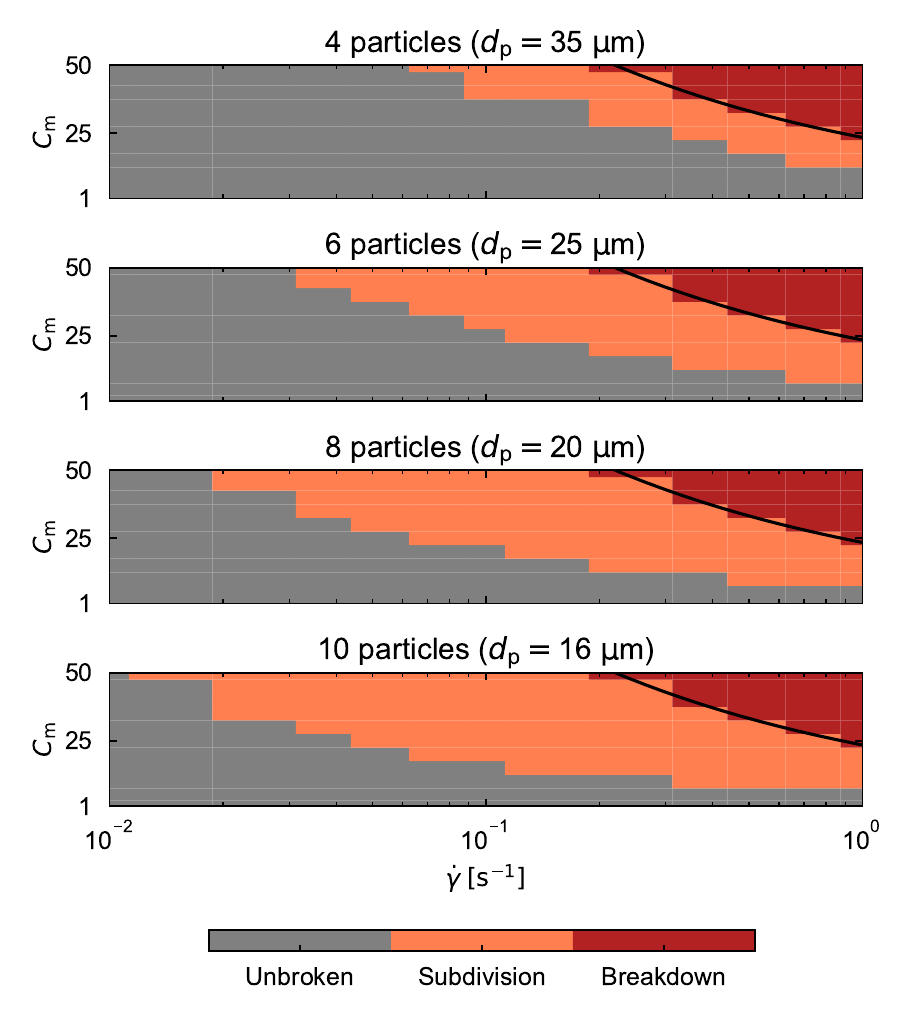}
\caption{Magnetised-particle chain behaviour under steady shear in a planar geometry with varying particle diameter (number of particles per chain), particle magnetisation (through the scaling constant $C_\mathrm{m}$) and shear rate ($\dot\gamma$). In black, chain complete-breakdown prediction through the Mason number defined by \citet{gans1999}: $Mn = 1$. Complete chain breakdown is independent of the number of composing particles and is well predicted by $Mn=1$, whereas larger chains are clearly more susceptible to subdivision (less stable).}
\label{fig:Chain_break_space}
\end{figure}

To have a clear idea of the possible effects of the geometrical changes on the particle chains, we wanted to set the system properties so that the chains remain unbroken in the perfectly parallel geometries while keeping the models relatively light. Therefore, we had an interplay between particle size, particle magnetisation and flow velocity. We performed simulations of a single chain on the parallel CP20 to test the limits of these characteristics. The particle diameter was varied between 35 and 16 \textmu m, leading to chains of 4 to 10 particles\footnote{The number of particles in each chain was calculated by rounding down (to the nearest integer) the required particle number to span the local gap height: $n = [h/d_\mathrm{p}]$, where $d_\mathrm{p}$ is the modelled particle diameter.} and the shear rate ranged from 0.01 to 1 s\textsuperscript{-1}, guaranteeing that the the particles are capable of effectively following the flow (for $d_\mathrm{p} = 35$ \textmu m, $St\lesssim3\times10^{-5}\ll1$). The magnetisation was progressively decreased using a factor, $1\leq C_\mathrm{m}\leq 50$, such that the tested magnetisation was: $M^* =M_\mathrm{sat.}/C_\mathrm{m}$ (with $M_\mathrm{sat.}=6.4$ Am\textsuperscript{2}/kg). From these simulations, three behaviours were observed: a) the chain remained unbroken, b) the chain undergoes subdivision (including combinations of minor chains and isolated particles), and c) the chain completely breaks down into individual particles. Examples of these behaviours are showcased in Figure \ref{fig:Chain_division_examples}, and the results can be seen in Figure \ref{fig:Chain_break_space}.

From Figure \ref{fig:Chain_break_space} we gather that chain breakdown is independent of the chain length and particle size, and is accurately predicted by a unitary Manson number as defined by \citet{gans1999} (adapting their Equation 23): 
\begin{equation}
    Mn = \frac{72\,\eta\,\dot\gamma}{\mu_0\,\mu_\mathrm{f}\,\rho_\mathrm{p}^2\,{M^*}^2}=1\,,
\end{equation}
which describes the ratio between viscous and magnetic effects (also plotted in Figure \ref{fig:Chain_break_space}). On the other hand, chain subdivision is clearly dependent on the chain length, being larger chains less stable\citep{gao2012}. We have yet to find a second Mason number that accurately predicts our chain subdivision, but this is out of the scope of this work.

From these results, we could select a shear rate and particle magnetisation that allowed unbroken chains in the parallel geometries, while minimising the computational effort. Similar to the experimental procedure, it is pertinent to have the same defined shear rate (rim shear rate, $\dot\gamma(R)$) for both geometries. While for PP geometries the shear rate varies along the radial axis (decreasing towards the centre), in CP geometries it remains constant (outside the truncated region); meaning the chains must remain unbroken under this defined shear rate. We opted for chains with only 4 particles ($d_\mathrm{p} = 35$ \textmu m) and a rim shear rate of $\dot\gamma(R)=1$ s\textsuperscript{-1}, having, thus, the maximum particle magnetisation factor before chain subdivision: $C_\mathrm{m} = 10$ (model particle magnetisation: $M^* = 0.64$ Am\textsuperscript{2}/kg).

Before heading to the results, it is worth mentioning that the characteristics of the geometries are significant to the chain dynamics. As was mentioned, the CP's constant shear rate along the radius corresponds to the maximum shear rate induced on the PP, therefore, for chains of equal length with an arbitrarily imposed rim shear rate, the CP is more likely to break the chains. On the other hand, the CP local gap height also depends on the radial position, which leads to smaller, more stable chains as we approach the centre. As such, an interplay between varying shear rate (on the PP) and chain length (on the CP) can diverge the magnetorheological results of either geometry.

\begin{figure}[htp]
\centering
\begin{minipage}[t]{\textwidth}
\centering
$\boldsymbol{\varphi = 0}$\\
\end{minipage}
\begin{minipage}[t]{.45\textwidth}
\centering
PP20
\end{minipage}
\begin{minipage}[t]{.45\textwidth}
\centering
CP20
\end{minipage}
\begin{minipage}[t]{\textwidth}
\centering
$t = 650$ s
\end{minipage}
\begin{minipage}[t]{0.45\textwidth}
\centering
\includegraphics[width=\linewidth]{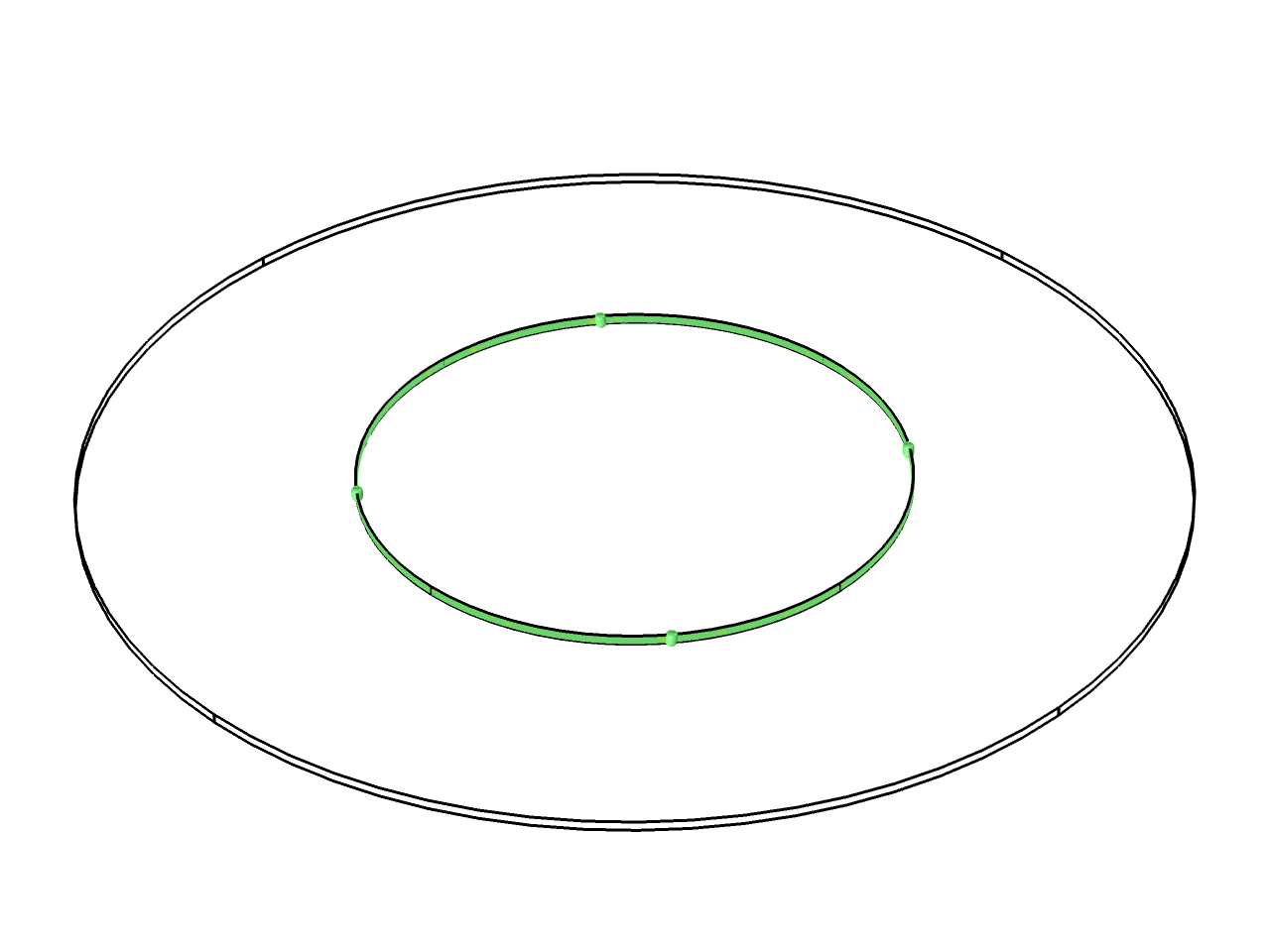}
\end{minipage}
\begin{minipage}[t]{0.45\textwidth}
\centering
\includegraphics[width=\linewidth]{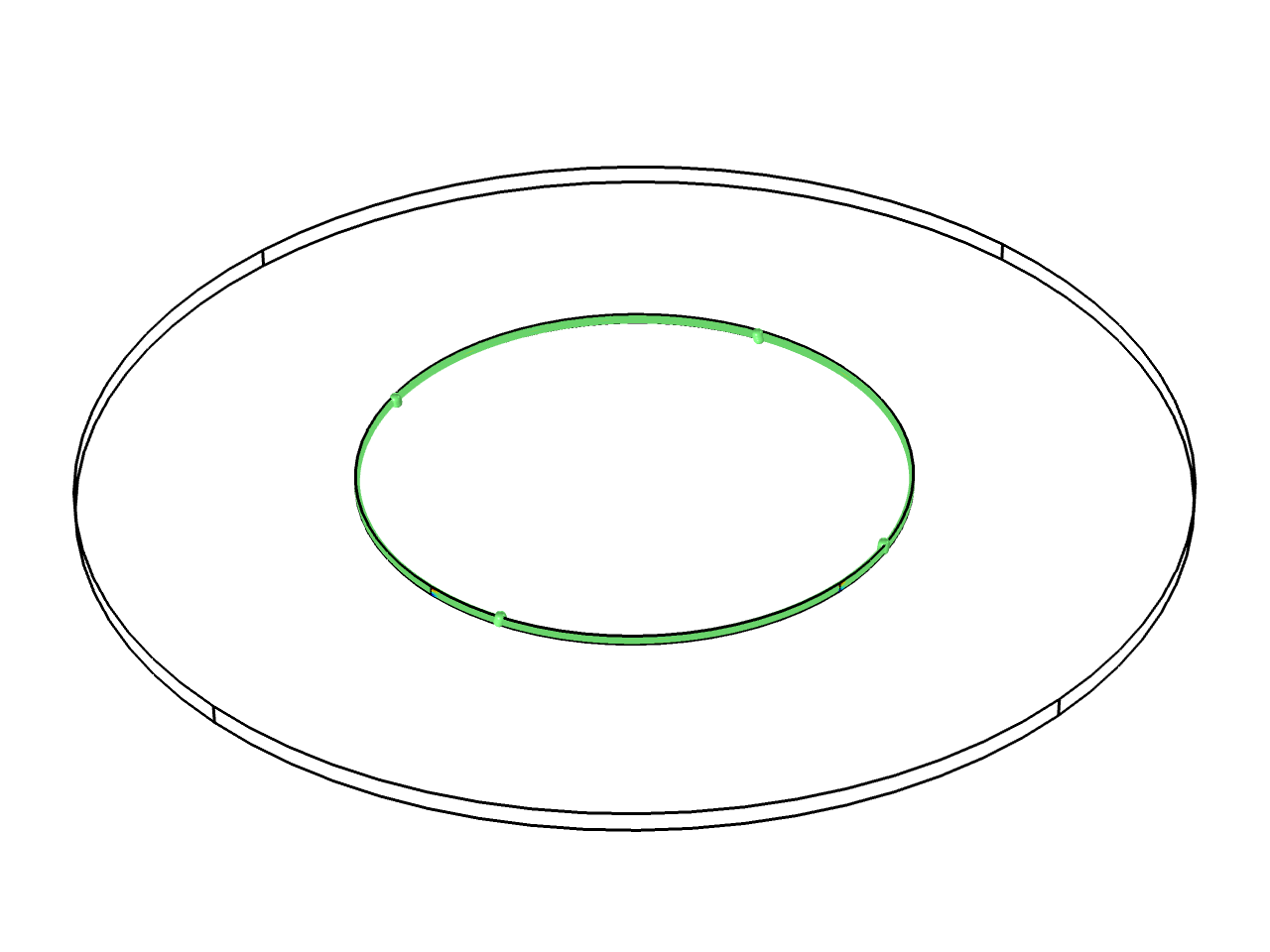}
\end{minipage}
\begin{minipage}[t]{\textwidth}
\centering
\includegraphics[width=\linewidth]{images/Numerical/Velocity_colorbar.pdf}
\end{minipage}
\caption{Magnetised-particle trajectories ($d_\mathrm{p} = 35$ \textmu m and $M^* = 0.64$ Am\textsuperscript{2}/kg) on the (left) PP20 and (right) CP20 models (without bottom plate inclination: $\varphi = 0$) with imposed shear rate $\dot\gamma = 1$ s\textsuperscript{-1} at time $t = 650$ s. In these perfectly-parallel geometries, the chains never subdivide and travel in a perfectly circular orbit.}
\label{fig:alpha0}
\end{figure}

In Figure \ref{fig:alpha0} are shown the particle trajectories in both geometries (without bottom plate inclination) at $t = 650$ s (the trajectories at other times, $t = 70$ and 200 s, are not included here for the sake of conciseness, but are provided as supplementary material). The chains remain unbroken, travelling along $r=R/2$ on either geometry, but moving faster on the CP20, where they are subject to a larger rotational velocity ($\Omega_\mathrm{CP} \approx 2\,\Omega_\mathrm{PP}$).

\begin{figure}[htp]
\centering
\begin{minipage}[t]{\textwidth}
\centering
$\boldsymbol{\varphi = 1^\circ}$\\
\end{minipage}
\begin{minipage}[t]{.45\textwidth}
\centering
PP20
\end{minipage}
\begin{minipage}[t]{.45\textwidth}
\centering
CP20
\end{minipage}
\begin{minipage}[t]{\textwidth}
\centering
$t = 70$ s
\end{minipage}
\begin{minipage}[t]{0.45\textwidth}
\centering
\includegraphics[width=\linewidth]{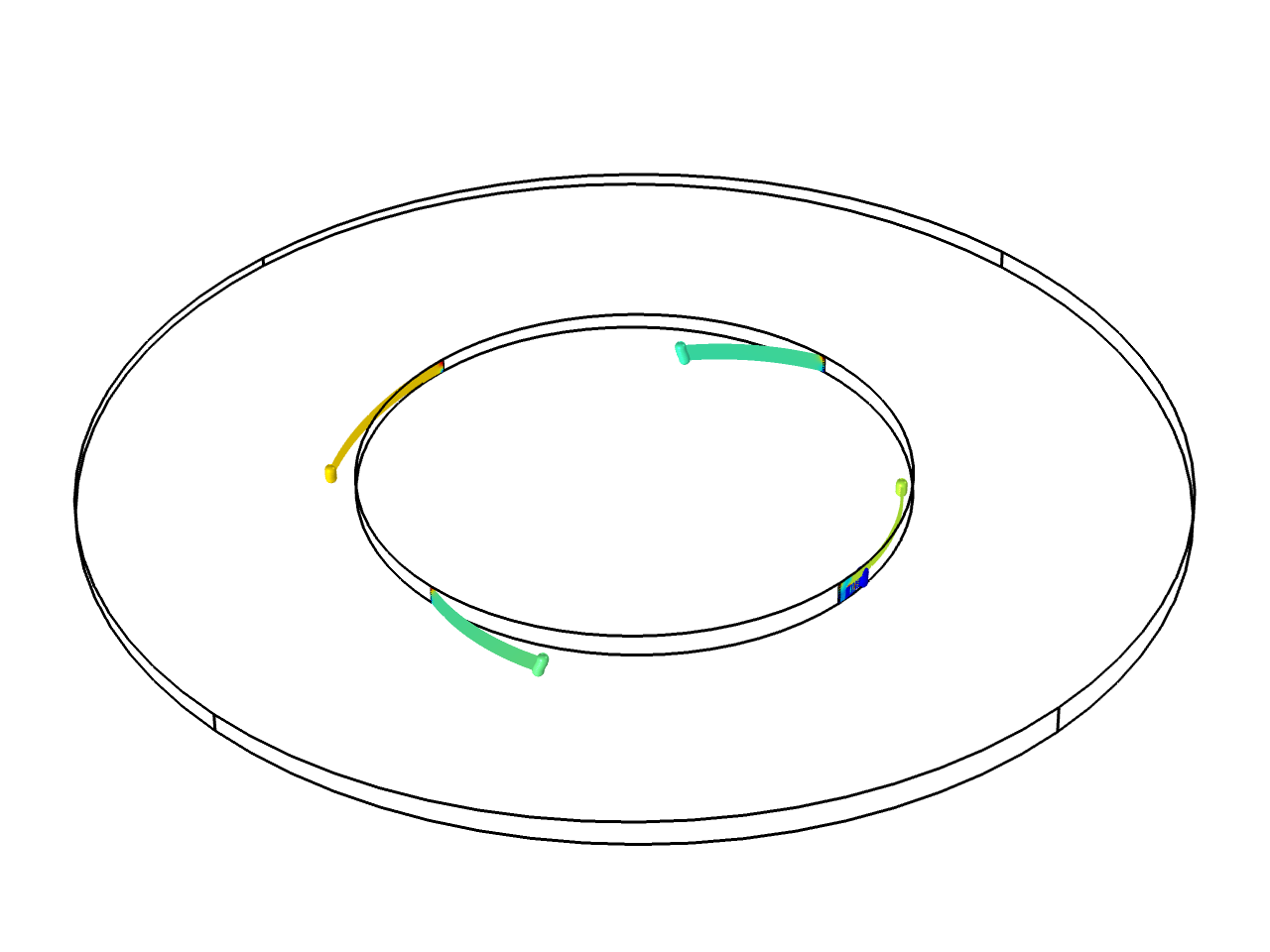}
\end{minipage}
\begin{minipage}[t]{0.45\textwidth}
\centering
\includegraphics[width=\linewidth]{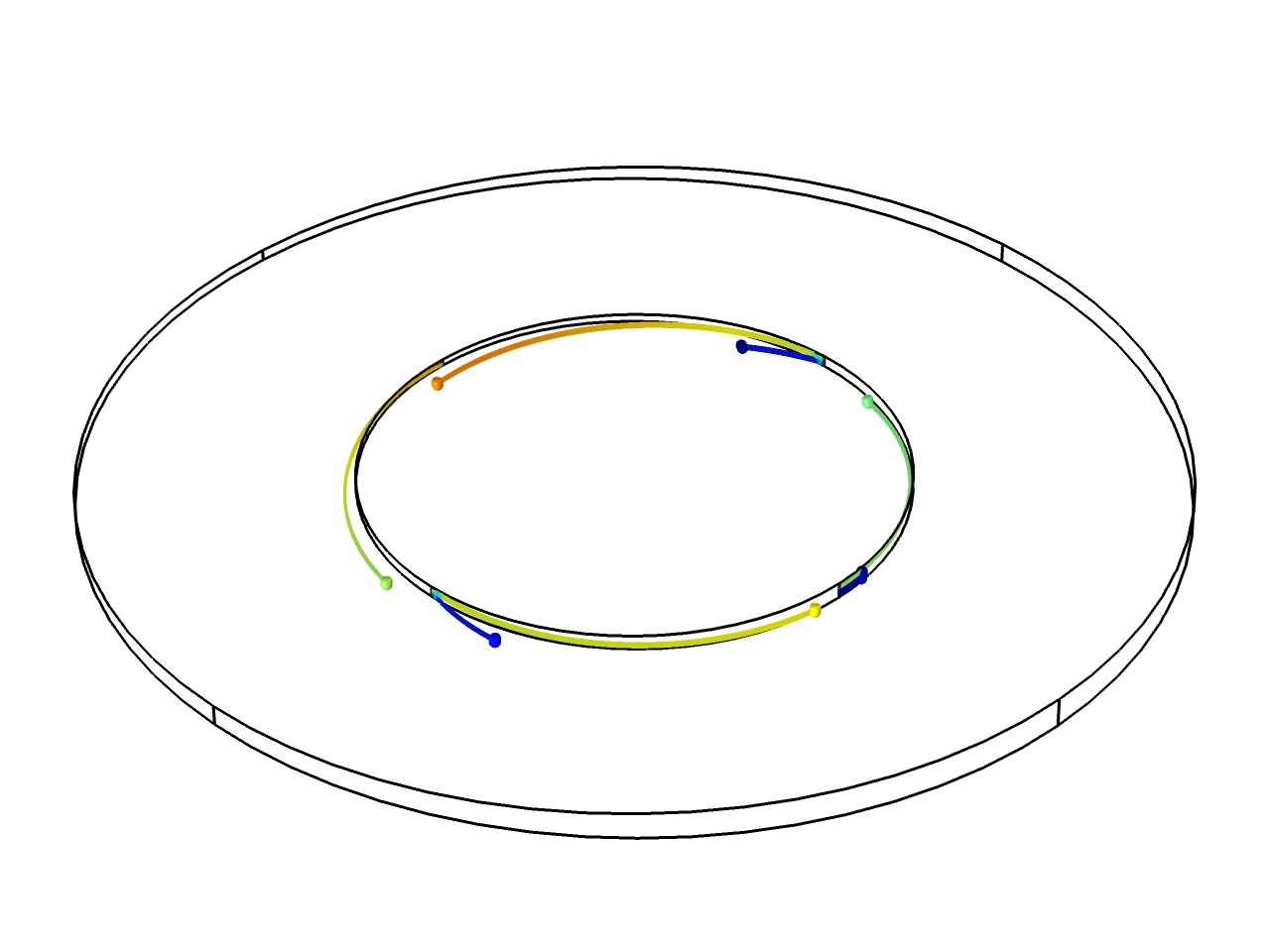}
\end{minipage}
\begin{minipage}[t]{\textwidth}
\centering
$t = 200$ s
\end{minipage}
\begin{minipage}[t]{0.45\textwidth}
\centering
\includegraphics[width=\linewidth]{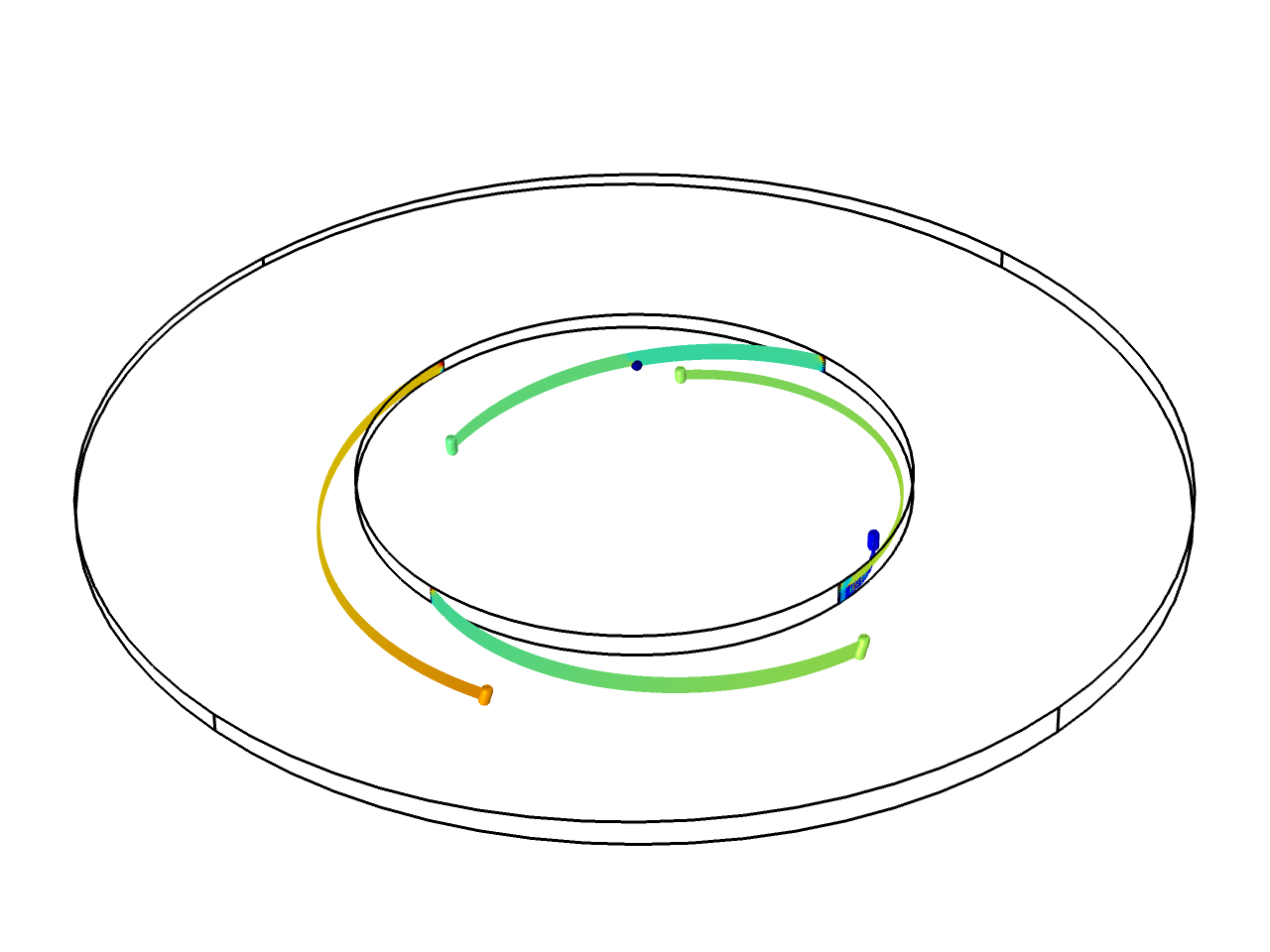}
\end{minipage}
\begin{minipage}[t]{0.45\textwidth}
\centering
\includegraphics[width=\linewidth]{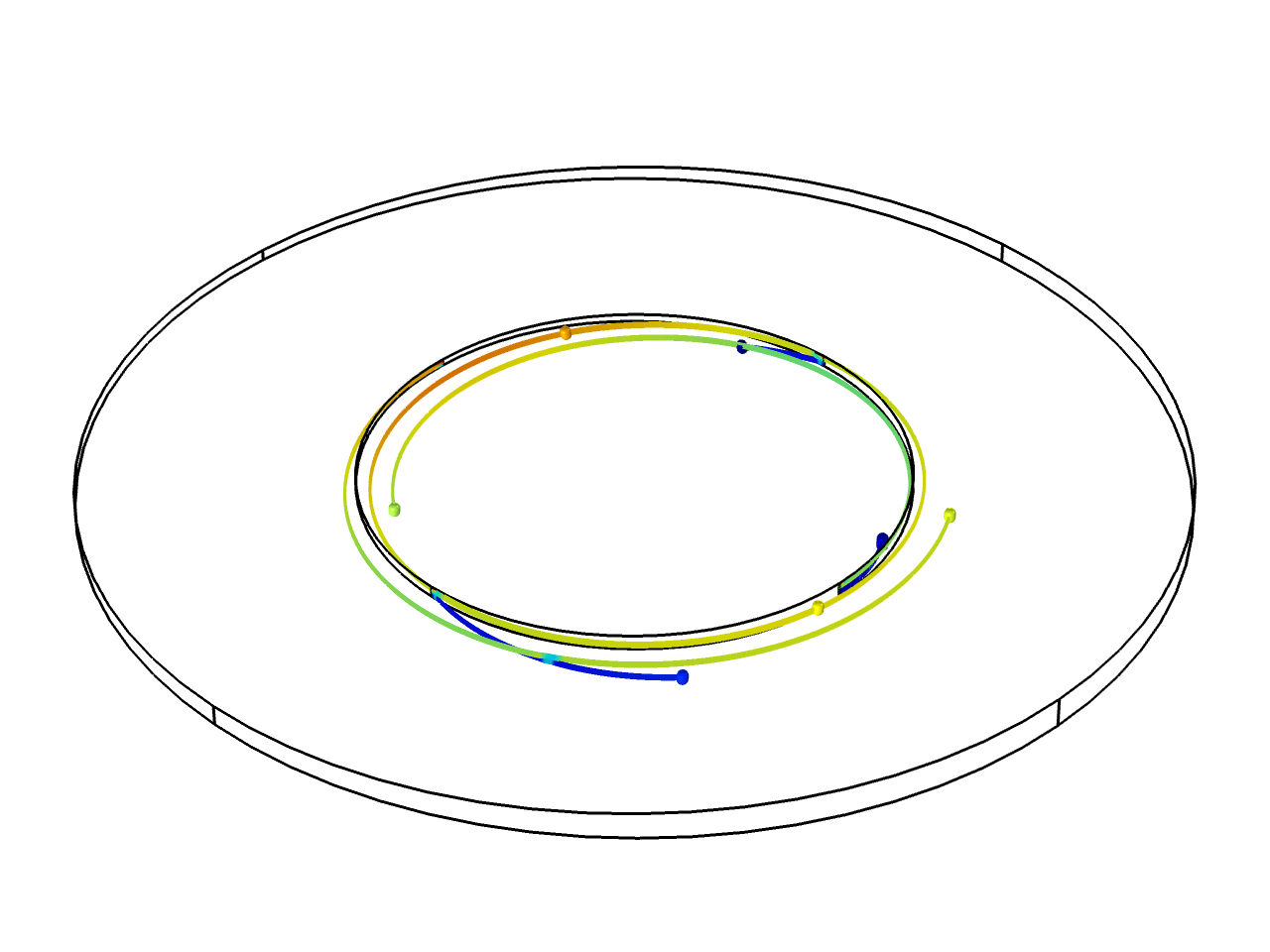}
\end{minipage}
\begin{minipage}[t]{\textwidth}
\centering
$t = 650$ s
\end{minipage}
\begin{minipage}[t]{0.45\textwidth}
\centering
\includegraphics[width=\linewidth]{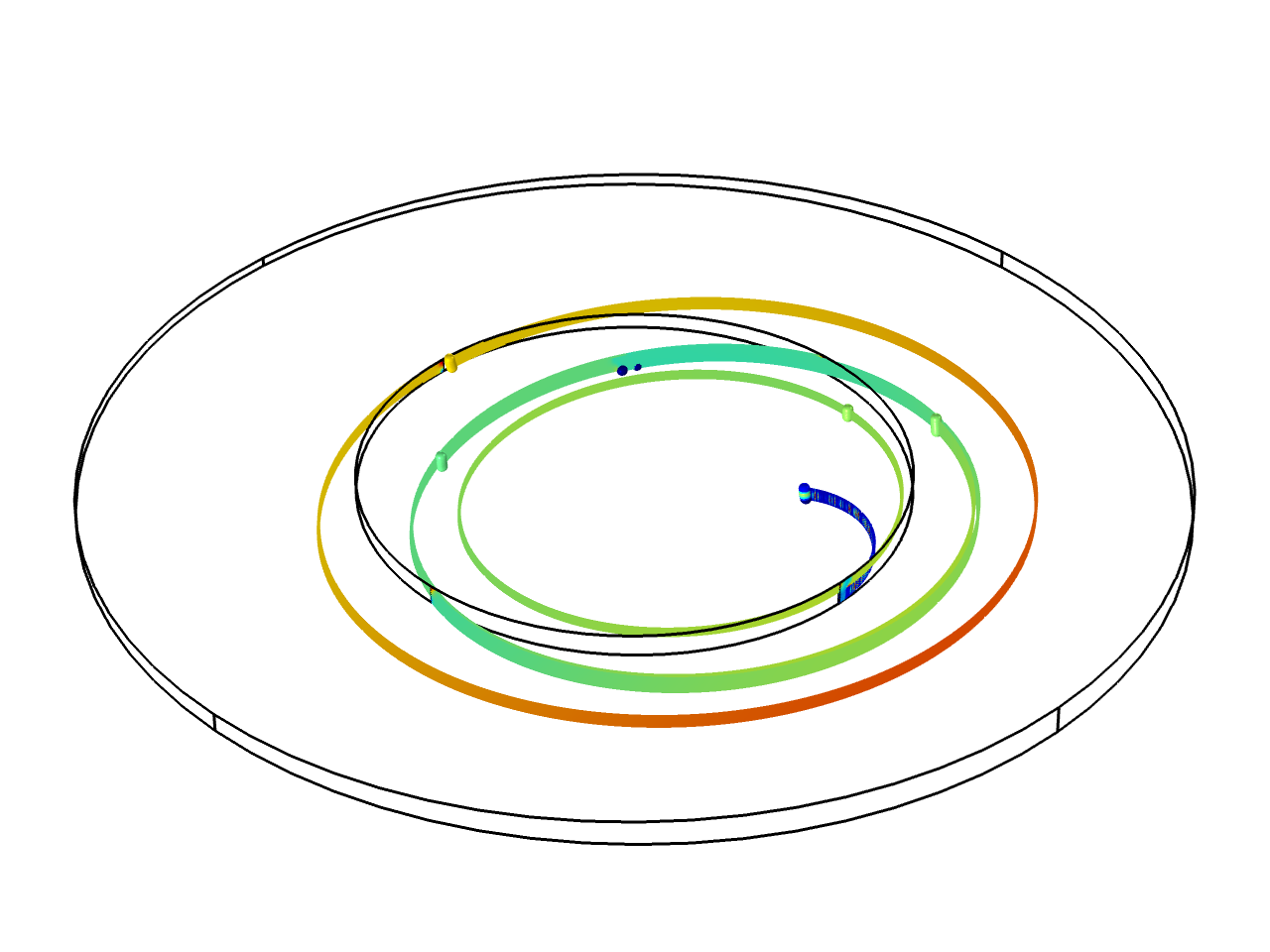}
\end{minipage}
\begin{minipage}[t]{0.45\textwidth}
\centering
\includegraphics[width=\linewidth]{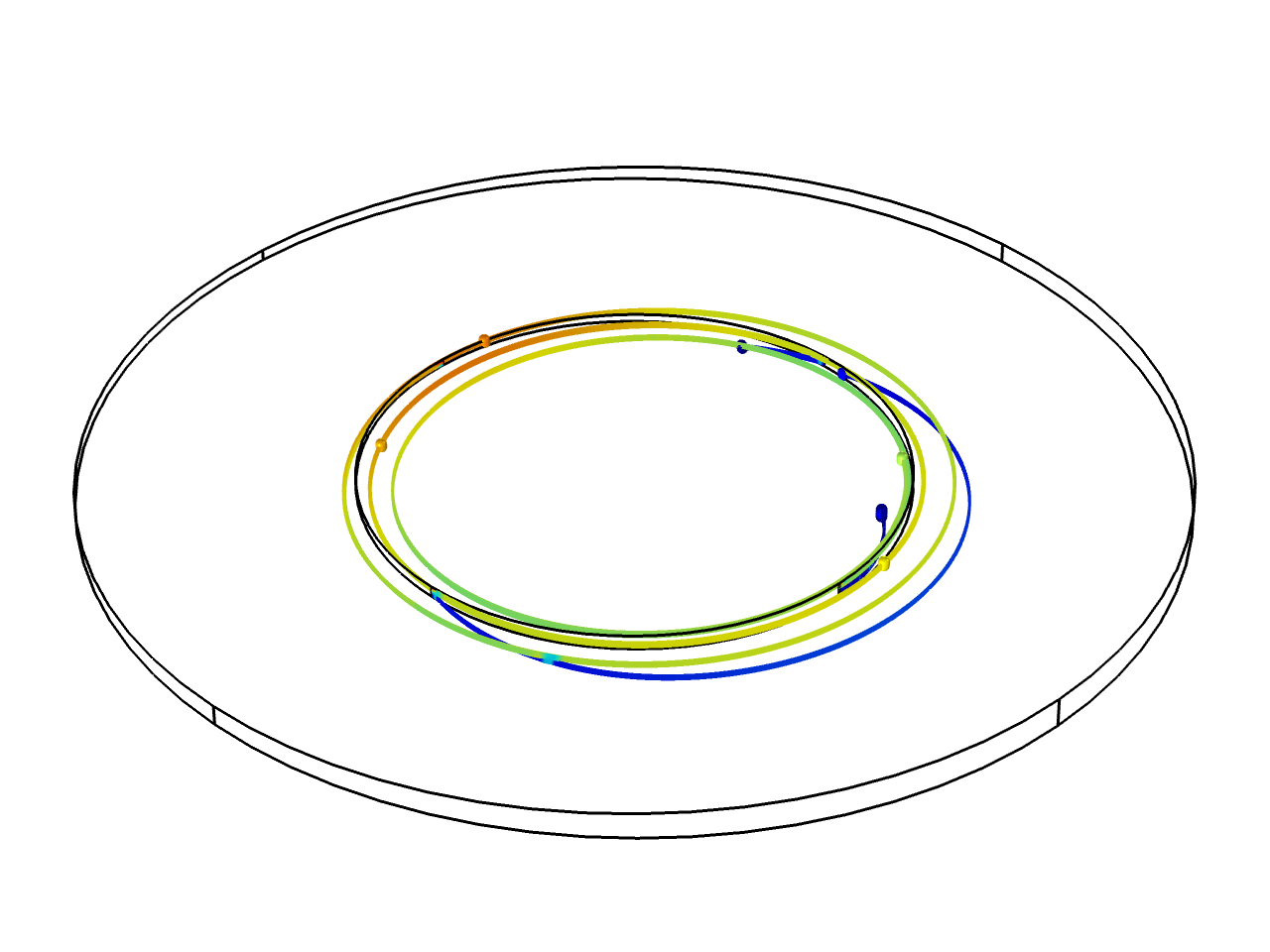}
\end{minipage}
\begin{minipage}[t]{\textwidth}
\centering
\includegraphics[width=\linewidth]{images/Numerical/Velocity_colorbar.pdf}
\end{minipage}
\caption{Magnetised-particle trajectories ($d_\mathrm{p} = 35$ \textmu m and $M^* = 0.64$ Am\textsuperscript{2}/kg) on the (left column) PP20 and (right column) CP20 models with a bottom plate inclination of $\varphi = 1^\circ$,  with imposed shear rate $\dot\gamma = 1$ s\textsuperscript{-1} at different times ($t = 70$, 200 and 650 s). Geometry non-parallelism leads to large alterations of the chain dynamics, inducing subdivision and more complex trajectories.}
\label{fig:alpha1}
\end{figure}

Introducing a bottom plate inclination of $\varphi=1^\circ$ (the maximum for acceptable viscosity measurements), the analogous results are shown in Figure \ref{fig:alpha1}. Before discussing the chain dynamics, the introduced geometry non-parallelism affects the number of particles in each chain. For the PP, it increases the chain length from 4 particles in each chain to 7, 9 and 12. For the CP, because the inclination is lesser than the cone angle ($\varphi < \beta$), the gap is reduced at the constrained region, $\theta = 0$, (see Figure \ref{fig:Models}) before increasing as we descend along the inclination, which leads to chains of 3, 6 and 8 particles.

Considering the PP results (left column of Figure \ref{fig:alpha1}), initially only the largest chain, of 12 elements, subdivides. Compared to the particle flow on the parallel geometry (Figure \ref{fig:alpha0}), the chains begin to diverge from the $R/2$ plane because the inclination of the geometry also leads to an inclination of the rotational axis. This becomes more evident as we move forward in time, particularly for $t = 650$ s, where the trajectories of most chains overlap. There are some additional observations regarding wall interactions. Because the intermediate chains (of 9 particles at $\theta = \pi/2$ and $3\pi/2$) do not subdivide, while travelling along the tilted geometry, they collide with the bottom wall and lose the particle that adheres to it. With a similar behaviour, the subdivided portion of the largest chain closest to the bottom wall also collides with it, but instead of further subdividing or losing the contact particle, the whole chain is stuck to the wall due to the weak local viscous effects. 

Considering the CP (right column of Figure \ref{fig:alpha1}), only the smaller chain of 3 elements remains unbroken. As the imposed rotational velocity is larger in this geometry, the enlarged shear rate leads to their subdivision despite the intermediate chains being three elements smaller than on the PP. In this case, because the intermediate chains are also broken, the three subdivided portions closest to the bottom wall stick to it when contact is made. Despite significantly altering the flow, this slight inclination does not reveal the counter-rotating region near the bottom wall. However, increasing the inclination angle to $\varphi = \beta = 1.981^\circ$ does generate a much stronger recirculation region (see Figure \ref{fig:Profiles - tilted}), which significantly affects the chain dynamics. For the sake of clarity, the particle trajectories for $\varphi = \beta$ are not shown here but are presented as supplementary material.

The bottom plate inclination calls for larger chains, which are easier to break, but the subdivided portions may be close in size to the original chains we ought to have in the perfectly parallel geometries. However, the issue with the magnetorheological measurements is not the chain size directly but the consequent chain velocity. Near the measuring geometry (top wall in this case), the closer the chain velocity is to the flow velocity, the lesser the influence on the measured torque and, therefore, the lesser the magnetic influence is felt. The particle trajectories shown in Figures \ref{fig:alpha0} and \ref{fig:alpha1} were plotted with a colour scheme in which the particle velocity is scaled with the maximum velocity at deployment ($\Omega\,R/2$), which allows us to see that the particles near the measuring plate/cone (on top) have a larger velocity on the tilted geometries than on the parallel ones, which will reduce the magnetically-induced viscosity increase. 

\subsection{Limitations}

Throughout this work we have considered a number of assumptions and simplifications that should be highlighted to contextualise the presented results. First, the non-parallelism hypothesis was brought forth as the most probable mechanism for the gathered experimental results but other geometrical characteristics and surface defects may be able to generate a similar response. For example, a small buckling-type deformation of the MP could also be responsible for a gap-error of similar magnitude and generate contraction/expansion flow on CP geometries. Nevertheless, we believe it is extremely unlikely that such surface defects would pass the supplier's quality control, which is also why we attribute the inclination of the MP to its mounting mechanism on the magnetorheological cell and not its surface itself.

Concerning the magnetorheological measurements, we have disregarded the gravitational effects, inertial issues and Brownian motion. Sedimentation should be counteracted by the magnetic dipole interactions up to a point, but the gravitational forces may still be relevant, particularly for weaker magnetic fields. The measurement time was kept short also to avoid this issue. We believe Brownian effects can be safely ignored, considering both the overpowering magnetic and viscous effects. Inertial issues, on the other hand, were disregarded on account of the individual particles' properties, but the larger aggregates that arise under the magnetic field influence may be affected, even though we find it unlikely to be significant.

The numerical work was conducted to simulate real steady shear measurements, but several simplifications were applied to reduced the computational effort. Experimentally, the sample/air interface is aimed to have a convex meniscus shape, dependent on the sample surface tension, and in this work it was considered as a straight wall with a free-slip boundary condition. Regarding the particle simulations, the magnetic field was considered perfectly uniform and the particles monodisperse. It was also considered that the particles that collide with the solid boundaries stick to it, when in reality the particle behaviour would depend on contact friction forces. The particle/fluid interactions were considered unilateral, where the particles have no effect on the flow field, which can alter the particle dynamics and the chain stability. This and the small number of simulated particles has limited our analysis to a qualitative evaluation of the chain dynamics, from which we extrapolate a larger-scale microstructural response and bulk rheological implications. More quantitative information could be gathered from a larger simulation, with multiphysics coupling and meaningful particle concentrations, but it requires considerable computational power, particularly because the flow asymmetry dictates a three-dimensional model.

\section{Conclusions}
\label{sec:conclusions}

In this work we set out to evaluate the suitability of an experimental setup for steady shear magnetorheological measurements of whole blood. An experimental campaign was conducted with Newtonian fluids in two planar geometries designed for magnetorheological measurements, one parallel-plates (PP20 MRD) and one cone-plate (CP20 MRD), and on two bottom plates, one for standard measurements (SP) and one for magnetic testing (MP).

It was found that the rheometer's minimum torque multiplied by a $70\times$ factor reasonably delimited low-shear errors and the apparent thickening at high shear was accurately predicted by the onset of secondary flows. On the MP we encountered a dependence of the PP-measured viscosity with gap height, returning lower viscosities as the gap is reduced, and a slight viscosity overestimation with the CP. The gap-error formulation\citep{kramer1987} effectively corrected the PP-measured viscosity data and the error itself was found to be around $\varepsilon\approx22.8$ \textmu m, most probably provoked by an inclination of the MP.

Numerical models of non-parallel geometries corroborated the experimental results, pointing towards an MP inclination between $0.10^\circ<\varphi<0.31^\circ$. The numerical work also revealed that the geometry non-parallelism can lead to notable flow alterations, particularly for CP geometries which essentially give rise to contraction/expansion flow and may lead to the emergence of counter-rotating regions as the inclination approaches the cone angle. We have found that for slight inclinations ($\lesssim1^\circ$) relatively small errors can be obtained ($\lesssim 10\%$) either by employing a CP or by correcting the PP data through the gap-error formulation. Nevertheless, we alert all users to the fact that geometry non-parallelism can have severe implications on the flow dynamics, particularly for CP geometries, and should be thoroughly evaluated as its own experimental error-source, instead of a mere gap-error mechanism. In the impossibility of a direct assessment of geometry parallelism, we recommend testing a CP geometry in addition to the usual PP gap-dependence study, as large inclinations can be identified through a CP-measured viscosity overestimation, at least with Newtonian fluids.

Magnetorheological measurements were also conducted with a Newtonian blood analogue seeded with magnetic particles. The results showed the typical magnetorheological response: a viscosity increase with particle concentration and magnetic field density. The flow of a few particle chains was modelled and it was observed that geometry non-parallelism can significantly affect the microstructural behaviour, weakening the strength of the magnetorheological response.

\section*{Supplementary material}

As supporting material we present: preliminary measurements, a description and discussion on the mesh employed on the numerical work, numerical estimations of the effects of a general gap-error on viscosity measurements with parallel geometries, and trajectories of magnetised particles in perfectly parallel geometries and with a bottom plate inclination equal to the angle of the employed cone-plate (obtained numerically).

\section*{Acknowledgements}

This work was financially supported by national funds through the FCT/MCTES (PIDDAC), under the project PTDC/EME-APL/3805/2021 (DOI 10.54499/PTDC/EME-APL/3805/2021), LA/P/0045/2020, UIDB/00532/2020 and UIDP/00532/2020, and the program Stimulus of Scientific Employment, Individual Support-2020.03203.CEECIND.

% references
\setcitestyle{numbers}
\bibliographystyle{unsrtnat}
\bibliography{main.bib}

\end{document}